\documentclass[aps,twocolumn,graphics,floatfix,tightenlines]{revtex4-2}
\usepackage{graphics}
\usepackage{epstopdf}
\usepackage{epsfig}
\usepackage{graphicx}
\usepackage{epsf,epic}
\usepackage{color}
\usepackage{caption}
\usepackage{subcaption}
\usepackage{amsmath}
\usepackage{booktabs}
\usepackage[flushleft]{threeparttable}
\usepackage{multirow}
\usepackage{physics}
\usepackage{hyperref}
\usepackage{amsfonts}
\usepackage{wrapfig}
\usepackage{breqn}
\usepackage{pstricks}
\usepackage[utf8x]{inputenc}
\usepackage{multirow}
\usepackage{fancyref}
\usepackage{pst-node}
\usepackage{float}
\usepackage{bm}
\usepackage{dcolumn}
\newcommand{\etal}{\textit{et al.\ }}
\newcommand{\ie}{\textit{i.e.\ }}

\makeatletter
\usepackage{etoolbox} % for \appto
\appto{\appendix}{%
	\@ifstar{\def\theequation@prefix{A.}}%
	{}%
}
\preto\maketitle{%
  \begingroup\lccode`~=`,
  \lowercase{\endgroup
  \let\saved@breqn@active@comma~% save breqn active comma
  \let~}\active@comma % set the active comma to what revtex4-1 wants
}
\appto\maketitle{%
  \begingroup\lccode`~=`,
  \lowercase{\endgroup
  \let~}\saved@breqn@active@comma % undo the change
}
\makeatother
\begin{document}
\title{Comparison of interband related optical transitions and excitons in ZnGeN$_2$ and GaN}
\author{Ozan Dernek}
\author{Walter R. L. Lambrecht}
\affiliation{Department of Physics, Case Western Reserve University,
  10900 Euclid Avenue, Cleveland, OH 44106}
\begin{abstract}
  The optical dielectric function of ZnGeN$_2$ is calculated from the interband
  transitions using the energy bands calculated in the quasiparticle self-consistent
  (QS)$G\hat W$ method using two different levels of approximation: the independent
  particle approximation (IPA) and the Bethe-Salpeter Equation (BSE) approach. The first
  allows us to relate peaks in $\varepsilon_2(\omega)$ to specific bands and
  {\bf k}-points but does not include electron-hole interaction effects. The second
  includes electron-hole interaction or excitonic effects. The corresponding changes in
  the shape of $\varepsilon_2(\omega)$ are found to be similar to those in GaN. The
  screened Coulomb interaction $\hat W$ is here calculated including electron-hole
  interactions in the polarization function and gives a band structure already going
  beyond the random phase approximation. The static dielectric constants including only
  electronic screening, commonly referred to as $\varepsilon^\infty$, were calculated
  separately by extrapolating the wave vector dependent macroscopic
  $\varepsilon_M({\bf q},\omega=0)$ for ${\bf q}\rightarrow0$. Below the quasiparticle
  gap, we find three bound excitons optically active for different polarization. The
  convergence of these bound excitons with respect to the density of the {\bf k}-mesh used
  in the BSE is studied and found to require a fine mesh. It is also found that these
  bound excitons originate from only the lowest conduction band and the top three valence
  bands. To incorporate the lattice screening, we include a scaling factor
  $(\varepsilon^\infty/\varepsilon^0)^2$, which allows us to obtain exciton binding
  energies of the correct order of magnitude similar to those in GaN. The excitons are
  related to each of the three fold split valence bands and the splittings of the latter
  are also studied as function of strain. Finally, a relation between the anisotropic
  effective masses and the valence band splitting is pointed out and explained. 
  \end{abstract}
\maketitle

\section{Introduction}
  ZnGeN$_2$ is a heterovalent ternary semiconductor closely related to wurtzite GaN. The
  $Pna2_1$ structure, which is the ground state structure, corresponds to a specific
  ordering of the divalent Zn and tetravalent Ge ions on the cation sublattice of
  wurtzite, in which each N is surrounded by two Zn and two Ge, thus locally obeying the
  octet rule. The resulting orthorhombic symmetry lowering leads to some distortion of the
  parent hexagonal lattice vectors away from the perfect $a=\sqrt{3}a_w$, $b=2a_w$ and
  $c=c_w$, with $a_w$ and $c_w$ are the corresponding wurtzite lattice constants. (Unlike
  in some of our previous papers on these materials, we here use the standard setting of
  the space group of the International Tables of X-ray Crystallography.) ZnGeN$_2$ and
  other heterovalent II-IV-N$_2$ ternaries have recently received increasing attention as
  being complementary to the well-studied III-N semiconductors, providing several
  opportunities for increased flexibility in tuning their properties. Several recent
  reviews are available on the current state of knowledge on crystal growth, electronic
  band structure, phonon, elastic, piezoelectric and defect properties of these
  materials \cite{Lambrechtbook,ictmc21,Martinez17}.

  Specifically for the Zn based Zn-IV-N$_2$, the electronic band structure was previously
  calculated \cite{Punya11} using the quasiparticle self-consistent $GW$ method, which is
  one of the most accurate and predictive methods available. It uses
  many-body-perturbation theory in the $GW$ approximation due to Hedin
  \cite{Hedin65,Hedin69} where $G$ is the one-electron Green's function and $W$ is the
  screened Coulomb interaction. The quasiparticle self-consistent version
  \cite{MvSQSGWprl,Kotani07} of this method, named QS$GW$, optimizes the non-interacting
  $H^0$, from which the dynamic self-energy effects are calculated as $\Sigma=iG^0W^0$, by
  extracting a static (\ie energy independent) and Hermitian ($\Re$) but non-local
  exchange-correlation potential
  $\tilde \Sigma_{ij}=\frac{1}{2}\Re{\{\Sigma_{ij}(\epsilon_i)+\Sigma_{ij}(\epsilon_j)\}}$.
  This self-energy $\tilde \Sigma_{ij}$ is obtained in the basis of eigenstates of $H^0$,
  and by iterating $H^0+\tilde \Sigma-v_{xc}$, with $v_{xc}$ the exchange-correlation
  potential in the initial $H^0$, to convergence. It thereby becomes independent of the
  starting $H^0$ which is usually taken as the density functional theory (DFT) Kohn-Sham
  Hamiltonian in either the local density approximation (LDA) or the generalized gradient
  approximation (GGA). The quasiparticle energies then become identical to the Kohn-Sham
  eigenvalues at convergence, hence the name ``quasiparticle'' self-consistent. However,
  the $H^0$ Hamiltonian is then no longer a Kohn-Sham Hamiltonian in the DFT sense, since
  it is not the functional derivative $\delta E^{DFT}/\delta n({\bf r})$ of a
  corresponding DFT total energy anymore.

  Here we revisit the QS$GW$ calculations of ZnGeN$_2$ for various reasons. First, there
  have been recent improvements in the QS$GW$ method including electron-hole effects in
  the screening of $W$. Second, the {\sc Questaal} codes \cite{questaal,questaalpaper}
  used for these calculations have been made more efficient, which allows us to check the
  convergence with stricter criteria. Third, the structural parameters used in previous
  calculations \cite{Punya11} were, in retrospect, not fully optimized. The main
  difference is that the larger $a/b$ ratio leads to a different ordering of the top two
  valence bands as will be discussed in Section \ref{sec:strain}. The structural
  parameters were already corrected in \cite{Quayle15}. To extract the best possible band
  structure and optical properties, we here use the experimental crystal structure
  parameters ($a = 5.45$ \AA, $b = 6.44$ \AA, and $c = 5.19$ \AA). The first focus of the
  paper is on calculating the optical dielectric function and analyzing it in terms of
  band-to-band transitions. To do this we calculate the optical dielectric function in the
  continuum region above the quasiparticle gap in the independent particle approximation
  (IPA) and including electron-hole interaction effects using the Bethe-Salpeter Equation
  (BSE) approach. We also perform similar calculations for the parent compound GaN to
  highlight the similarities in optical properties and to facilitate the comparison by
  considering the band structure folding effects from the GaN wurtzite Brillouin zone to
  the $Pna2_1$ Brillouin zone.
  
  Our second aim is to understand the valence band maximum splitting and the resulting
  exciton fine-structure below the band-to-band continuum. We study the convergence of the
  exciton gaps with increasingly finer {\bf k}-mesh near $\Gamma$ and show how their
  polarization is related to the symmetry of the valence bands involved and study their 
  dependence on strain. Finally, we point out a relation of the symmetry of the top three
  valence bands to the anisotropy of the effective mass tensor of each of these bands.

\section{Computational Method}
  As already mentioned, the calculations in this study are carried out using the
  {\sc Questaal} code suite \cite{questaalpaper}, which implements the density functional
  theory (DFT) using a full-potential linearized muffin-tin-orbital (FP-LMTO) basis set
  and extends beyond DFT to many-body-perturbation theory by incorporating the $GW$ and
  BSE methods. For details about the QS$GW$ method, see Ref. \onlinecite{Kotani07} and for
  a full description of the FP-LMTO method, see Ref. \onlinecite{questaalpaper}. While the
  QS$GW$ method is a great advance compared to DFT calculations in terms of the accuracy
  of band gaps and other band structure features, it tends to somewhat overestimate the
  band gaps because it underestimates the screening by not incorporating the
  electron-hole interactions in the screening of the screened Coulomb interaction
  $W=(1-Pv)^{-1}v$, where $v$ is the bare Coulomb interaction and $P$ is the two-point
  polarization propagator. This is commonly known as the random phase approximation
  (RPA). Calculations using an exchange-correlation kernel $f_{xc}$, extracted from BSE
  calculations or using a bootstrap kernel approximation
  \cite{Shishkin07,ChenPasquarello15} within time-dependent DFT (TDDFT),
  \ie $W=[1-P(v+f_{xc})]^{-1}v$, have shown that this tends to reduce the self-energy by a
  factor of $\sim$0.8 and this has lead to the commonly used approach, namely the
  $0.8\Sigma$ approximation \cite{Deguchi16,Chantis06a,Bhandari18}. That approach was also
  used in the previous calculations on ZnGeN$_2$ \cite{Punya11,Quayle15}.

  Recently, a new approach, calculating directly the four-point polarization function via
  a BSE at all ${\bf q}$-points and then recontracting it to two-point function was
  introduced by Cunningham \etal \cite{Cunningham18,Cunningham21,Radha-LCO21}. This
  approach is equivalent to adding a vertex correction to the polarization function in the
  Hedin set of equations with the vertex based on the
  $\delta\Sigma^{GW}(12)/\delta G(34)$. However, it was argued that within the spirit of
  the QS$GW$ approach, no corresponding vertex corrections are needed in the self-energy
  because of the cancellation of the $Z$-factor between the coherent part of the Green's
  function and the vertex. The $Z$-factor $Z=(1-d\Sigma/d\omega)^{-1}$ measures the
  reduction of the coherent quasiparticle part of the dynamic Green's function $G$
  compared to the independent particle Green's function $G^0$. Since only $G^0$ is made
  self-consistent in the QS$GW$ method, not $G$ itself, the $Z$-factor is omitted. In any
  case, this new approach, named QS$G\hat W$, removes the arbitrariness of an {\sl ad-hoc}
  correction factor.

  While in the previous calculations of ZnGeN$_2$, the 0.8$\Sigma$ approach was used, we
  here use the QS$G\hat W$ approach. Secondly, other factors still limit the accuracy of
  our previous calculations of ZnGeN$_2$. The calculations of \cite{Punya11} used, in
  retrospect, a somewhat imperfectly relaxed crystal structure. This was later corrected
  in \cite{Quayle15} where we used a GGA relaxed structure within the
  Perdew-Burke-Ernzerhof parametrization. Here, we prefer to use the actual experimental
  structure because the typical overestimate of the lattice constants, in particular the
  volume per unit cell, leads to an underestimation of the gap.

  The present calculations also use a larger basis set than in the previous work
  \cite{Punya11,Quayle15}. LMTO envelope functions are included with two smoothed Hankel
  function basis sets including up to $l\le3$ and $l\le2$ respectively and standard
  choices of the smoothing radii and Hankel function kinetic energy $\kappa^2$ suitable
  for $GW$ calculations. In addition, the Zn $3d$ orbitals are treated as augmented
  orbitals while the $4d$ are included as local orbital to better represent the high-lying
  conduction band states. For Ge, the $3d$ semi-core orbitals are included as local
  orbital while the $4d$ are augmented orbitals. Augmentation inside the spheres is done
  up to $l\le4$. By augmentation we mean that the expansion in spherical harmonics of a
  basis envelope function centered on one site about another site is replaced by a
  combination of the solutions of the radial Schr\"odinger equation
  $\phi_l(\epsilon_\nu,r)$ at a linearization energy $\epsilon_\nu$ and its energy
  derivative $\dot{\phi}_l$ that matches the envelope in value and slope. This expansion
  is carried out by means of structure constants. All calculations are including scalar
  relativistic effects in the potential. In the $GW$ self-energy calculations, we include
  all the bands generated by this basis set. This includes bands up to about 10 Ry. Two
  point quantities are expanded in an auxiliary basis set of products of partial waves
  inside the spheres and products of interstitial plane waves. This basis set describes
  screening as embodied in the dielectric response function
  $\varepsilon^{-1}_{IJ}({\bf q},\omega)$ more efficiently than a plane wave expansion and
  thereby bypasses the need for higher conduction bands in the generation of the
  dielectric function. Here the subscripts $I,J$ label the auxiliary basis set. The
  short-wavelength screening is taken care of by the products of partial waves inside the
  spheres. Other convergence parameters include the {\bf k}-mesh on which the self-energy
  is calculated for which we went up to $4\times4\times4$ and found negligible difference
  from the $3\times3\times3$ mesh. Using the atom-centered LMTO basis set the self-energy
  can then be expanded in real space and transformed back by Fourier transformation to a
  finer {\bf k}-mesh used for the self-consistent charge density or spectral functions,
  and for the {\bf k}-points along symmetry lines. In other words, the LMTO atom centered
  basis set serves as a Wannier-type expansion for interpolations with the difference that
  it is not an orthogonal basis set and thereby more localized. The self-energy matrix in
  the basis set of the Kohn-Sham eigenstates is fully included instead of only carrying
  out a perturbative calculation of the Kohn-Sham to $GW$ eigenvalue shifts. To facilitate
  the above mentioned interpolation of the self-energy, which requires sufficiently well
  localized basis functions, it approximates the full self-energy matrix by a diagonal
  average value above a certain energy cut-off chosen to be 2.71 Ry here. This average
  value of the self-energy is evaluated over a range of 0.5 Ry above this cut-off. The
  rationale behind these approximations and other details about the method are explained
  in Ref. \cite{Kotani07}
  
  The main focus of the present paper is the optical interband transition properties. We
  calculate the imaginary part of the dielectric function $\varepsilon_2(\omega)$ at
  various levels of approximation. First, in the IPA one uses
    \begin{eqnarray}
      \varepsilon_2(\omega)&=&\frac{8\pi^2e^2}{\Omega \omega^2}\sum_v\sum_c\sum_{{\bf k} \in BZ}f_{v{\bf k}}(1-f_{c{\bf k}}) \nonumber \\
      &&|\langle \psi_{{v\bf k}}|[H,{\bf r}]|\psi_{c{\bf k}}\rangle|^2\delta(\omega-\epsilon_{c{\bf k}}+\epsilon_{v{\bf k}}). \label{eq:AdlerWiser}
    \end{eqnarray}
  Here, the dielectric function is calculated as a sum over band-to-band transitions
  weighted by the matrix elements of the velocity operator. In this calculation we used a
  fine and well-converged $16\times16\times16$ {\bf k}-mesh. In case of the bands being
  calculated at the QS$GW$ or QS$G\hat W$ level, the non-local exchange correlation
  potential leads to additional terms beyond the momentum matrix elements, which involve
  $d\tilde \Sigma/dk$. The advantage of this approach is that, in principle, it takes the
  ${\bf q}\rightarrow0$ limit exactly. However, technical difficulties in evaluating
  $d\tilde{\Sigma}/dk$ lead to some overestimate of the optical matrix elements.
  Alternatively, one can evaluate directly $\varepsilon_2({\bf q},\omega)$, which would
  involve matrix elements of $e^{i{\bf q}\cdot{\bf r}}$ rather than the commutator, at
  small but finite ${\bf q}$ and take the limit numerically by extrapolation.

  The main advantage of the sum over transitions expression of Eq.~\ref{eq:AdlerWiser} is
  that we can take the optical function apart into its separate band-to-band
  contributions. Comparing these with the interband differences plotted as functions of
  {\bf k} along symmetry lines, we can identify how peaks in the $\varepsilon_2(\omega)$
  contributions are related to singularities (corresponding to critical points) or flat
  regions of the interband differences (corresponding to parallel bands) and therefore
  high joint-density of states (JDOS) contributions.

  On the other hand, this approach does not include electron-hole interaction or local
  field effects which modify the shape of the optical dielectric function. These can be
  obtained within the BSE approach to many-body-perturbation theory. Here, the key step is
  to calculate the four-point polarization 
    \begin{equation}
      \bar{P}(1234)=P^0(1234)+\int d(5678)P^0(1256)K(5678)\bar{P}(7834), \label{eq:BSE1}
    \end{equation}
  with the  kernel,
    \begin{equation}
      K(1234)=\delta(12)(34)\bar{v}-\delta(13)\delta(24)W(12). \label{eq:kernel}
    \end{equation}
  where $\bar{v}_{{\bf G}}({\bf q})=4\pi/|{\bf q}+{\bf G}|^2$ if ${\bf G}\ne0$, and zero
  otherwise, is the microscopic part of the bare Coulomb interaction and we have expressed
  it here in terms of a plane wave basis set. The macroscopic dielectric function is then
  given by
    \begin{equation}
      \varepsilon_M(\omega)=1-\lim_{{\bf q}\rightarrow0}v_{{\bf G}=0}({\bf q})\bar{P}_{{\bf G}={\bf G}^\prime=0}({\bf q},\omega). \label{eqepsmac}
    \end{equation}
  The contribution from $\bar{v}$ adds local-field effects while the contribution from
  $W$, or in our case $\hat W$ leads to the explicit excitonic effects on the dielectric
  function. Here, the four-point polarization depends on the coordinates, spin and time of
  four particles. The numbers $1,2,3,4$ are a short hand for position, spin and time of
  each particle, in other words $1\equiv({\bf r}_1,\sigma_1,t_1)$. In practice, this
  integral equation is solved by introducing a resonant transition state basis set
  $\psi_{v{\bf k}}({\bf r}_h)^*\psi_{c{\bf k}}({\bf r}_e)$, which are products of
  one-particle eigenstates. Specifically, 
    \begin{eqnarray}
      P({\bf r},\bar{{\bf r}},{\bf r}^\prime,\bar{{\bf r}^\prime})&=&\sum_{n_1,n_2,n_3,n_4}P_{n_1n_2n_3n_4}\nonumber \\ &&\psi_{n_1}({\bf r})^*\psi_{n_2}(\bar{{\bf r}})\psi_{n_3}({\bf r}^\prime)\psi_{n_4}(\bar{{\bf r}}^\prime)^*.
    \end{eqnarray}
  The polarization is then given by
    \begin{equation}
      P_{n_1n_2n_3n_4}(\omega)=(f_{n_4}-f_{n_3})(H^{2p}-\omega)^{-1},
    \end{equation}
  with the two-particle Hamiltonian
    \begin{equation}
      H^{2p}_{n_1n_2n_3n_4}=(\epsilon_{n_1}-\epsilon_{n_2})\delta_{n_1n_3}\delta_{n_2n_4}-K_{n_1n_2n_3n_4}(\omega),
    \end{equation}
  where $K$ uses a similar expansion of $K(1,2,3,4)$ in the basis functions of the
  one-particle Hamiltonian. In the present work a static approximation $\omega=0$ for the
  $W$, hence for the kernel $K$, and the Tamm-Dancoff approximation of only keeping
  resonant and not anti-resonant transitions are used. In the above, the basis state
  indices $n_1$, $n_3$ can then be identified with valence bands $v{\bf k}$,
  $v^\prime{\bf k}^\prime$ respectively and $n_2,n_4$ with conduction band states
  $c{\bf k}$, $c^\prime{\bf k}^\prime$. In the first term only direct vertical transition
  $\epsilon_{c{\bf k}}-\epsilon_{v{\bf k}}$ occur, while the kernel term mixes states at
  different ${\bf k}$ and ${\bf k}^\prime$, as well as between different valence and
  conduction band pairs. In other words, the two-particle Hamiltonian of the BSE equation
  $H_{vc{\bf k},v^\prime c^\prime{\bf k}^\prime}$ has dimensions $N_vN_cN_k$ with $N_v$
  the number of valence bands, $N_c$ the number of conduction bands and $N_{\bf k}$ the
  number of ${\bf k}$ points included. The ${\bf k}$-points are taken on a regular mesh in
  the Brillouin zone. The equations used here and approximations follow the review paper
  of Onida \etal \cite{Onida02}. Diagonalizing this two-particle Hamiltonian yields the
  eigenvalues and eigenvectors of the excitons in terms of the one-particle eigenstates,
  from which the dielectric function can be obtained. The same cautions apply about the
  accuracy of the optical dipole matrix elements as mentioned above. The eigenstates of
  the two-particle Hamiltonian, however, are mixtures of different vertical transitions
  and hence the decomposition in individual band-to-band transitions, strictly speaking,
  no longer makes sense. We should also mention that we here only calculate the spin
  singlet excitons. As explained in \cite{Rohlfing2000}, the spin structure, if no
  spin-polarization is present and we ignore spin-orbit coupling, can readily be taken
  into account. The singlet excitons then involve $2\bar{v}-W$ in the kernel in
  Eq.~\ref{eq:kernel}. For the calculations of the full energy range of
  $\varepsilon_2(\omega)$ we choose $N_v=24$, $N_c=16$ which encompasses all N-$2p$
  derived valence bands and conduction bands up to a similar energy above the conduction
  band minimum. To examine the excitons in more detail we use a smaller set of $N_v=3$ and
  $N_c=1$ which were found to be the main contributing bands to these excitons but with a
  larger {\bf k}-mesh which is required to converge the exciton binding energies. 

\section{Results}
\subsection{Band structure}\label{sec:band}

    \begin{figure}
      \includegraphics[width=9cm]{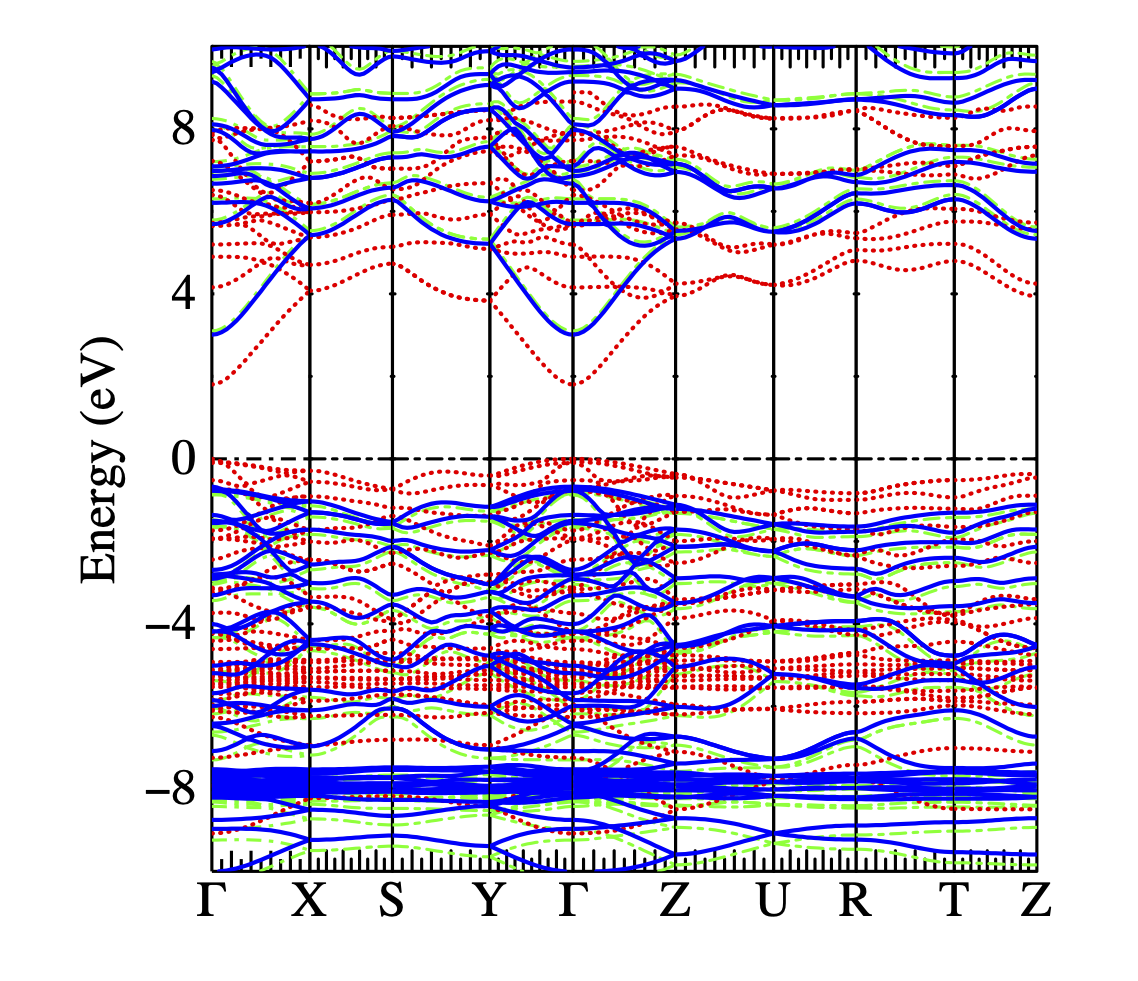}
      \caption{Band structure of $Pna2_1$ ZnGeN$_2$ at experimental lattice parameters in
               GGA (red dotted), QS$GW$ (green dashed) and QS$G\hat W$ (blue solid). The
               zero is placed at the VBM of GGA.\label{figbnds}}
    \end{figure}

  The band structure of ZnGeN$_2$ is shown in Fig.~\ref{figbnds} in three different
  approaches: GGA, QS$GW$ and QS$G\hat W$. The zero of energy is placed at the valence
  band maximum (VBM) of the GGA. One can thus see that in QS$GW$, the VBM shifts down by
  0.77 eV while the conduction band minimum (CBM) shifts up by 1.30 eV. Adding the ladder
  diagrams reduces the self-energy shifts of both band edges by a small amount. The band
  gaps are summarized in Table~\ref{tabgaps} and for comparison, we include the
  corresponding results for GaN. For both QS$GW$ and QS$G\hat W$ calculations we used
  $4\times4\times4$ {\bf k}-mesh for ZnGe$N_2$. To be consistent, an $8\times8\times4$
  {\bf k}-mesh was used for GaN. One can see that the self-energy correction to the gap
  $\Delta \Sigma_{QSGW}= E_g(QSGW)-E_g(GGA)$ and
  $\Delta \Sigma_{QSG\hat W}= E_g(QSG\hat W)-E_g(GGA)$ have a ratio of
  $\Delta \Sigma_{QSG\hat W}/\Delta \Sigma_{QSGW}=0.89$ in ZnGeN$_2$, and the same ratio
  is 0.88 for GaN. Thus, the inclusion of ladder diagrams reduces the self-energy gap
  correction by slightly larger than 10 \%, which is somewhat smaller than the often used
  20 \% reduction in the 0.8$\Sigma$ approach. We caution that this ratio is sensitive to
  the number of bands $N_v$ and $N_c$ in the QS$G\hat{W}$ calculations. We here used
  $N_v=24$ valence and $N_c=16$ conduction bands for ZnGeN$_2$. To obtain the same
  accuracy, we use $N_v=6$ and $N_c=4$ for GaN. As seen in Table~\ref{tabgaps}, using a
  higher $N_c=10$ reduces the gap slightly more, so that the QS$GW$ to QS$G\hat{W}$ gap
  reduction factor becomes 0.81 instead of 0.88. However, this would then also lead to a
  smaller $\hat{W}$ in the BSE calculations, which would correspondingly reduce the
  exciton binding energies and shifts the peaks between IPA and BSE, but the actual
  exciton gaps would stay similar. Furthermore, the gaps depend strongly on the lattice
  constants. For ZnGeN$_2$, we had used experimental lattice constants but for GaN, the
  value of $a=3.215$ \AA\ corresponds to the calculated PBE-GGA value and is slightly
  overestimated. Using the room temperature experimental lattice constant of 3.189 \AA\ 
  \cite{Maruska69,Leszcynski96}, we obtain a higher gap by 0.23 eV. We should note that
  these gaps do not include zero-point-motion (ZPM) corrections or finite temperature
  corrections. The ZPM band gap renormalization is expected to be of order $-$0.15 eV in
  ZnGeN$_2$ by comparison with GaN \cite{Nery16}.

    \begin{table}[h]
      \caption{Band gaps in eV for ZnGeN$_2$, and GaN in different approximations. The GaN
               band gap is reported for two different lattice constants and calculations
               where we include different number of conduction bands in the QS$G\hat{W}$
               approach.\label{tabgaps}}
      \begin{ruledtabular}
        \begin{tabular}{lcccc}
                               &                                & GGA                    & QS$GW$                  & QS$G\hat W$       \\ \hline
          ZnGeN$_2$            &                                & 1.801                  & 3.874                   & 3.641             \\ \hline \hline
          \multirow{4}{*}{GaN} & \multirow{2}{*}{$a=3.189$ \AA} & \multirow{2}{*}{1.884} & \multirow{2}{*}{3.868}  & 3.624$^{\ast}$    \\
                               &                                &                        &                         & 3.487$^{\dagger}$ \\ \cline{2-5} 
                               & \multirow{2}{*}{$a=3.215$ \AA} & \multirow{2}{*}{1.697} & \multirow{2}{*}{3.638}  & 3.392$^{\ast}$    \\
                               &                                &                        &                         & 3.261$^{\dagger}$ \\
        \end{tabular}
      \end{ruledtabular}
        \begin{tablenotes}
          \small
            \item $^{\ast}$ $\hat W$ was calculated using $N_c=4$.
            \item $^{\dagger}$ $\hat W$ was calculated using $N_c=10$.
        \end{tablenotes}
    \end{table}

    \begin{figure}
      \includegraphics[width=6cm]{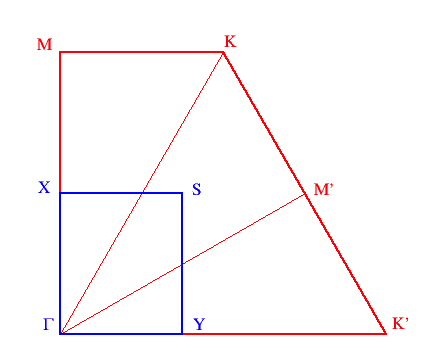}
        \caption{Relation of hexagonal to $Pna2_1$ Brillouin zone.\label{figbzs}}
    \end{figure}
    \begin{figure}
      \includegraphics[width=9cm]{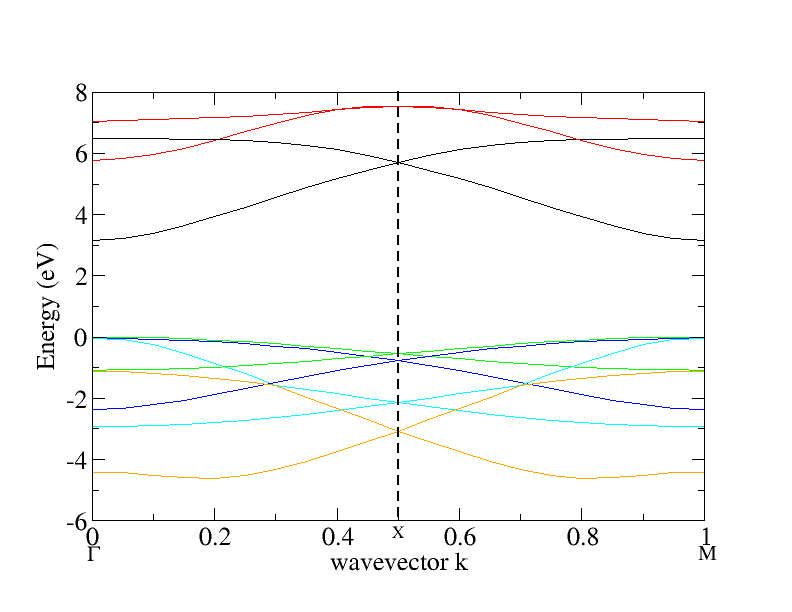}
        \caption{Folding of GaN bands along wurtzite $\Gamma-M$ onto $\Gamma-X$ line of
                 $Pna2_1$.\label{figGM}}
    \end{figure}

    \begin{figure}
      \begin{subfigure}[b]{0.5\textwidth}
        \centering
          \includegraphics[width=8cm]{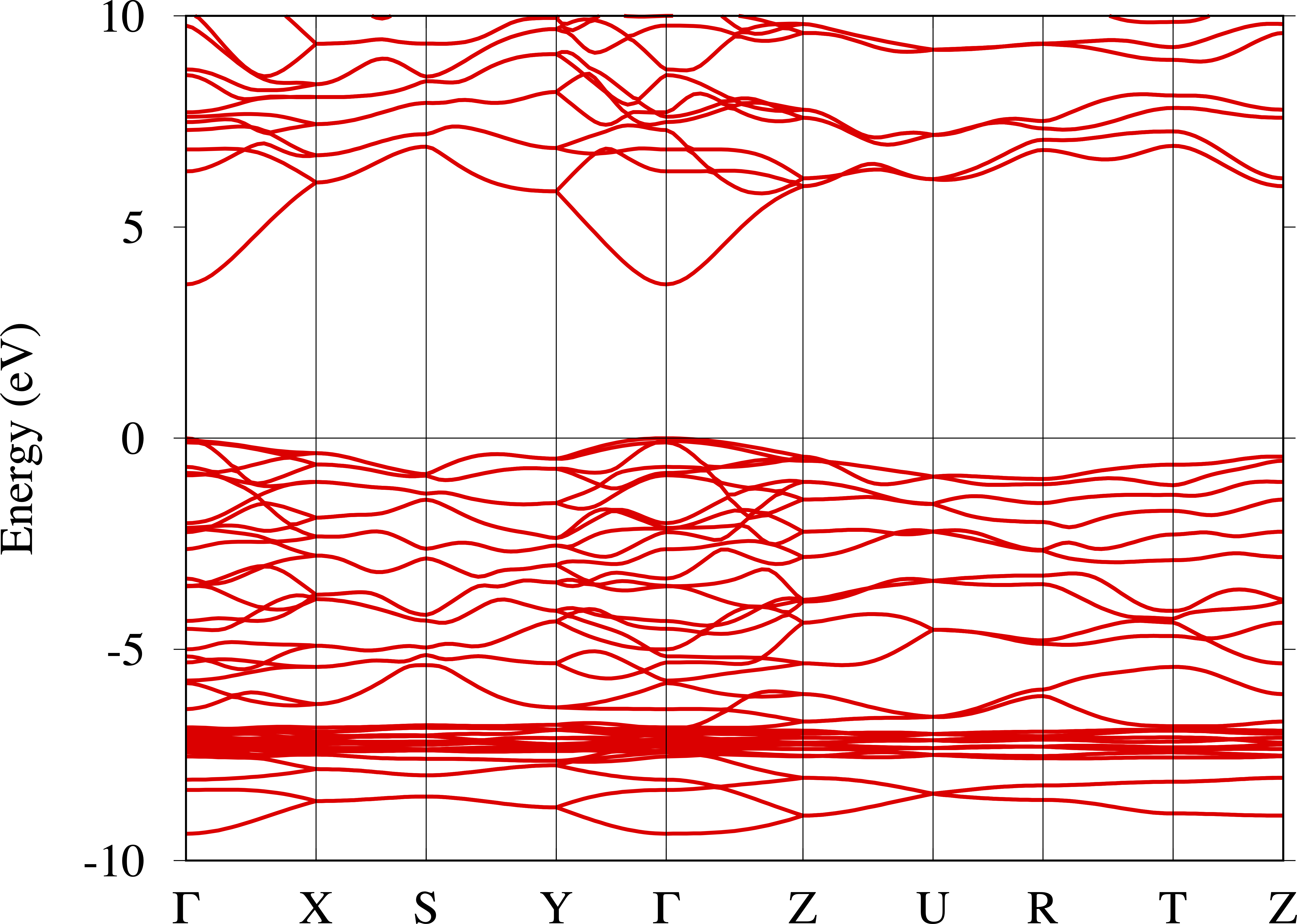}
          \label{}
      \end{subfigure}
        \hfill
      \begin{subfigure}[b]{0.5\textwidth}
        \centering
          \includegraphics[width=8cm]{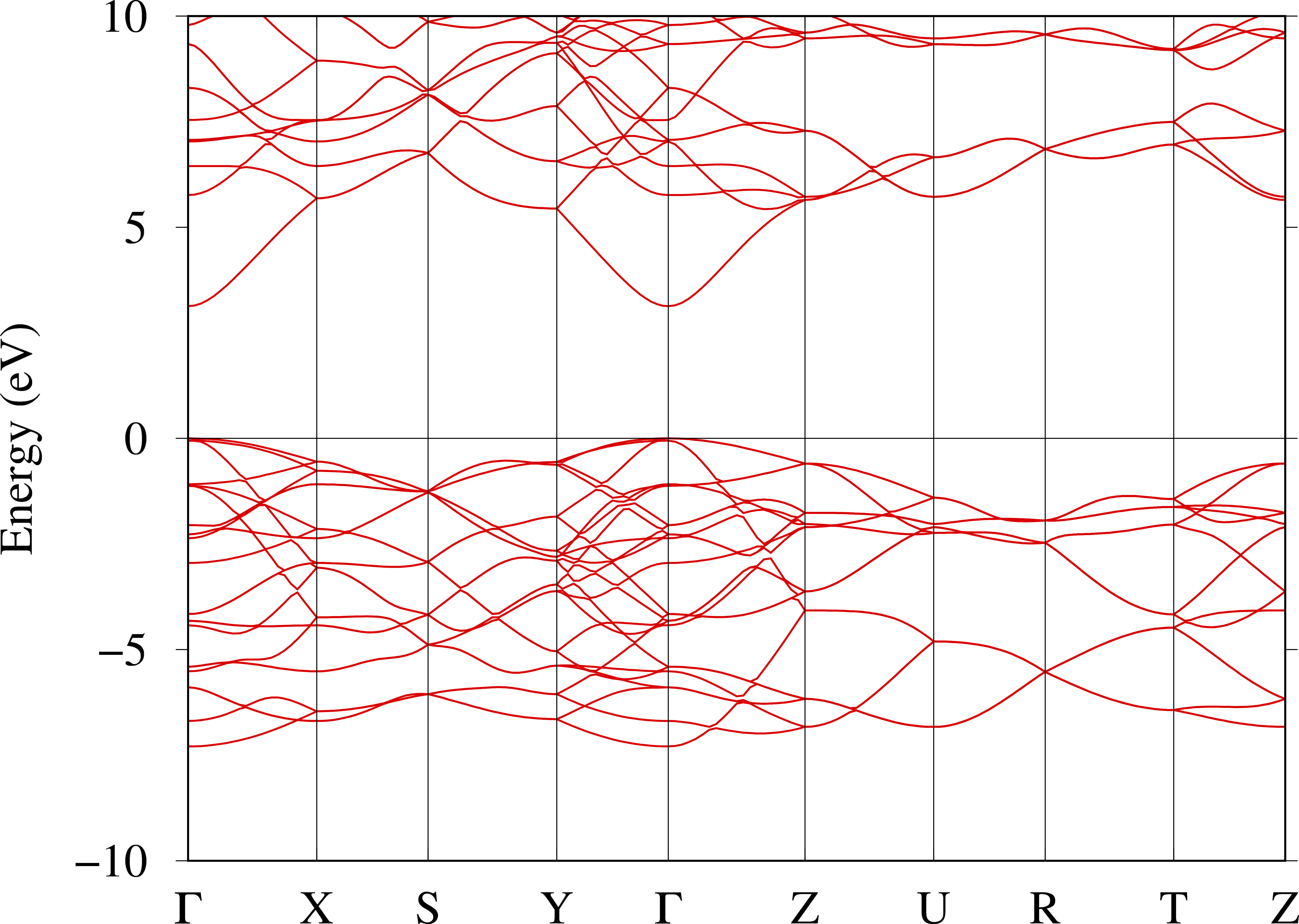}
          \label{}
      \end{subfigure}
        \caption{Band structure of ZnGenN$_2$ (top) compared to GaN (bottom) both in
                 $Pna2_1$ structure.\label{figbndzgngan}}
    \end{figure}
      
  It is instructive to compare the band structure of ZnGeN$_2$ with that of its parent
  compound GaN. The band structure of GaN in the wurtzite structure is well known and is
  shown in Supplementary Material \cite{supmat} in the usual hexagonal Brillouin zone as
  obtained with the present computational method. Here we discuss how it relates to the
  ZnGeN$_2$ band structure by band folding. 
  
  The $Pna2_1$ structure can be viewed as $2\times2$ supercell of the wurtzite, with
  ${\bf a}^o=2{\bf a}_2^w+{\bf a}_1^w$, ${\bf b}^o=2{\bf a}_1^w$ and
  ${\bf c}^o={\bf c}^w$. The relation of the $Pna2_1$ Brillouin zone to that of wurtzite
  is illustrated in Fig.~\ref{figbzs}. One can see that the $M-K$ line will be folded
  about the $X-S$ line onto the bottom $\Gamma-K'$ line and subsequently a second folding
  takes place about the $Y-S$ line. Thus the eigenvalues of the $M$ point will now occur
  at $\Gamma$ but also the $M'$ point is folded onto $X$, so the same $M$-eigenvalues will
  also occur at $X$. In Fig.~\ref{figGM} we show how the bands along $\Gamma-M$ of
  wurtzite are found back in folded fashion along $\Gamma-X$ by a single folding about the
  point $X$. Additional details of the band folding of $K$ are discussed in Supplementary
  Material \cite{supmat}.

  These help to understand the band structure of the GaN 16 atom supercell directly
  plotted in the $Pna2_1$ Brillouin zone, which can then in turn be compared with that of
  ZnGeN$_2$ as shown in Fig.~\ref{figbndzgngan}. The correspondence is seen to be quite
  close, especially in the conduction band region. The perturbation due to the Zn-Ge
  difference in potential is thus rather small. Of course, in the deeper valence band
  region near $-8$ eV we recognize the close set of bands from the Zn $3d$ bands which in
  Ga lie much deeper and are not included in the presently shown energy range. The results
  for GaN in the supercell were obtained from those in the primitive wurtzite cell by
  converting the self-energy matrix to real space using the procedure of \cite{Dernek22}
  followed by a simple band structure step without having to reconstruct the
  self-consistent potential. Just as $M-L$ is folded onto $\Gamma-X$ the points above it
  along the $z$ direction $L$ will fold onto $Z$ of $Pna2_1$. The whole $M-L$ line is thus
  folded onto $\Gamma-Z$. This will be seen to be relevant in the next section. 
  
\subsection{Dielectric function in BSE}\label{sec:BSE}
    \begin{figure}
      \includegraphics[width=8cm]{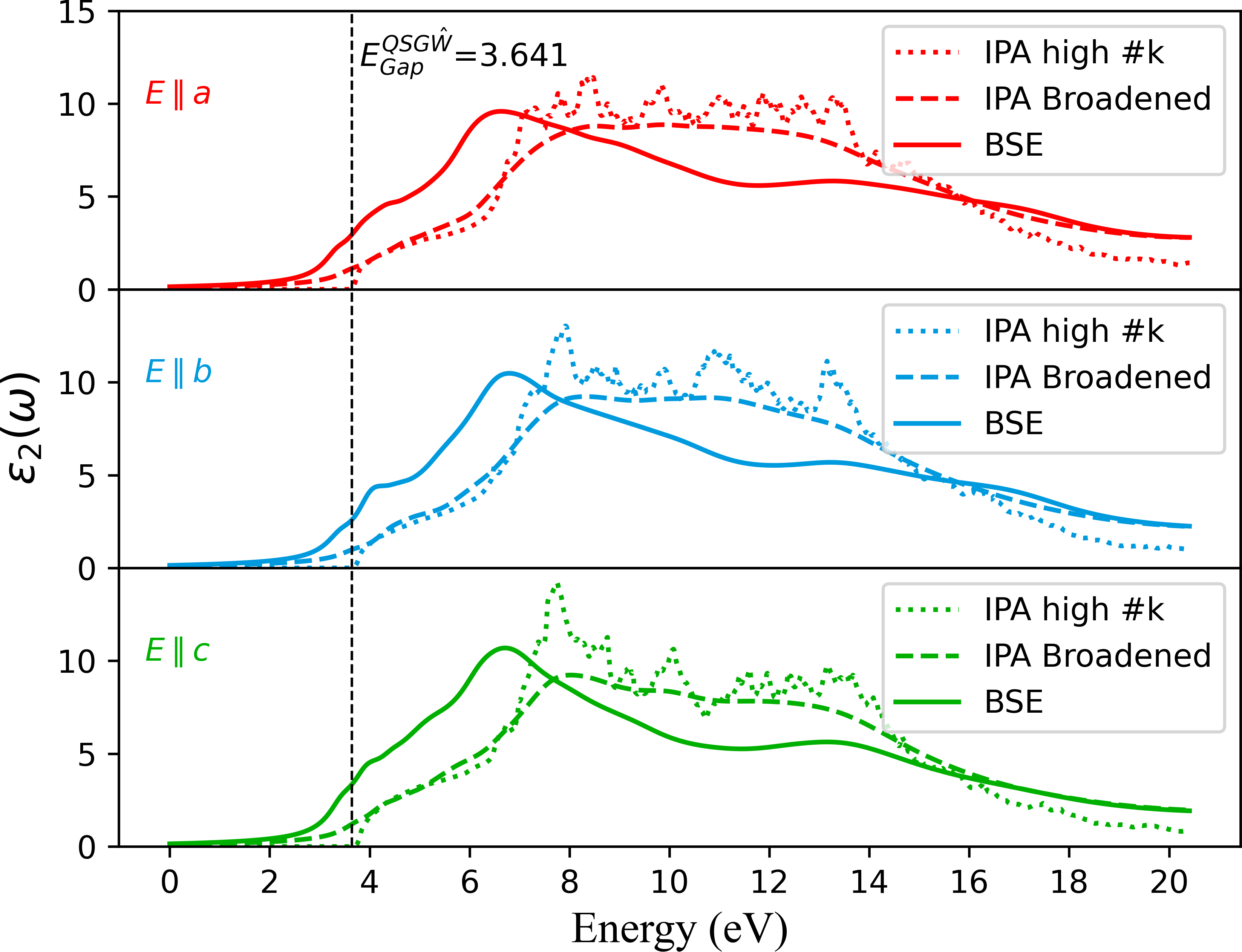}
        \caption{Imaginary part of the dielectric function of ZnGeN$_2$ for three
                 polarizations and comparing IPA with BSE approximations. Imaginary part
                 of energy is set to 0.02 Ry. The IPA is once calculated via
                 Eq.~\ref{eq:AdlerWiser} with a $16\times16\times16$ {\bf k}-mesh and the
                 tetrahedron integration method, and also with an imaginary part of
                 $\omega$ and a $6\times6\times6$ mesh. The latter {\bf-k}-mesh was also
                 used in BSE calculation.\label{figepsbse}}
    \end{figure}

  In Fig.~\ref{figepsbse} we compare the dielectric function in the IPA with the BSE
  calculation for ZnGeN$_2$. We show the IPA result once with a converged number of
  {\bf k}-points as obtained from Eq.~\ref{eq:AdlerWiser} and once with the same
  broadening and {\bf k}-point mesh as used in the BSE, where we can only afford a less
  dense {\bf k}-mesh. The broadening is chosen to match the results of the IPA as closely
  as possible between the two approaches and amounts to an imaginary part of the energies
  of $\sim$0.3 eV. For a smaller broadening, unphysical oscillations appear in the
  spectrum. These oscillations result from the coarse sampling which misses the
  eigenvalues at points in between the sampled {\bf k}-points and make it difficult to
  discern actual peak structure from artifacts of the {\bf k}-point sampling. 

  The overall spectrum of $\varepsilon_2(\omega)$ in Fig.~\ref{figepsbse} can be described
  as follows. First, there is a region of slowly increasing $\varepsilon_2(\omega)$
  ($\propto\sqrt{E-E_{onset}}$) between 3.6 eV and about 6 eV in both BSE and IPA. In the
  BSE at 6 eV a new feature sets in, peaking at about 6.5 eV, and decreasing to about 10
  eV followed by another broad peak between about 12 and 14 eV after which the intensity
  decreases. In the IPA, the first peak reaches its maximum at a somewhat higher energy,
  around 7.5 eV but the spectrum does not show a minimum near 10 eV and instead shows a
  broad series of peaks of about the same value till 14 eV. One can see that the BSE
  overall shifts oscillator strength to lower energies and results in a clearer two broad
  peaks structure: one between 6 and 10 eV and one between 10 and 16 eV. The first peak in
  BSE appears to occur at the onset of the steep rise in $\varepsilon_2(\omega)$ of the
  IPA spectrum. The quasiparticle gap is indicated by the vertical dashed line. With a
  smaller imaginary part, the BSE also shows peaks below the quasiparticle gap, which
  correspond to bound excitons but with the broadening in Fig.~\ref{figepsbse} these
  features are smeared out and appear only as a shoulder below the quasiparticle gap. We
  will discuss the excitons separately in a later section.

\begin{figure}
      \includegraphics[width=8cm]{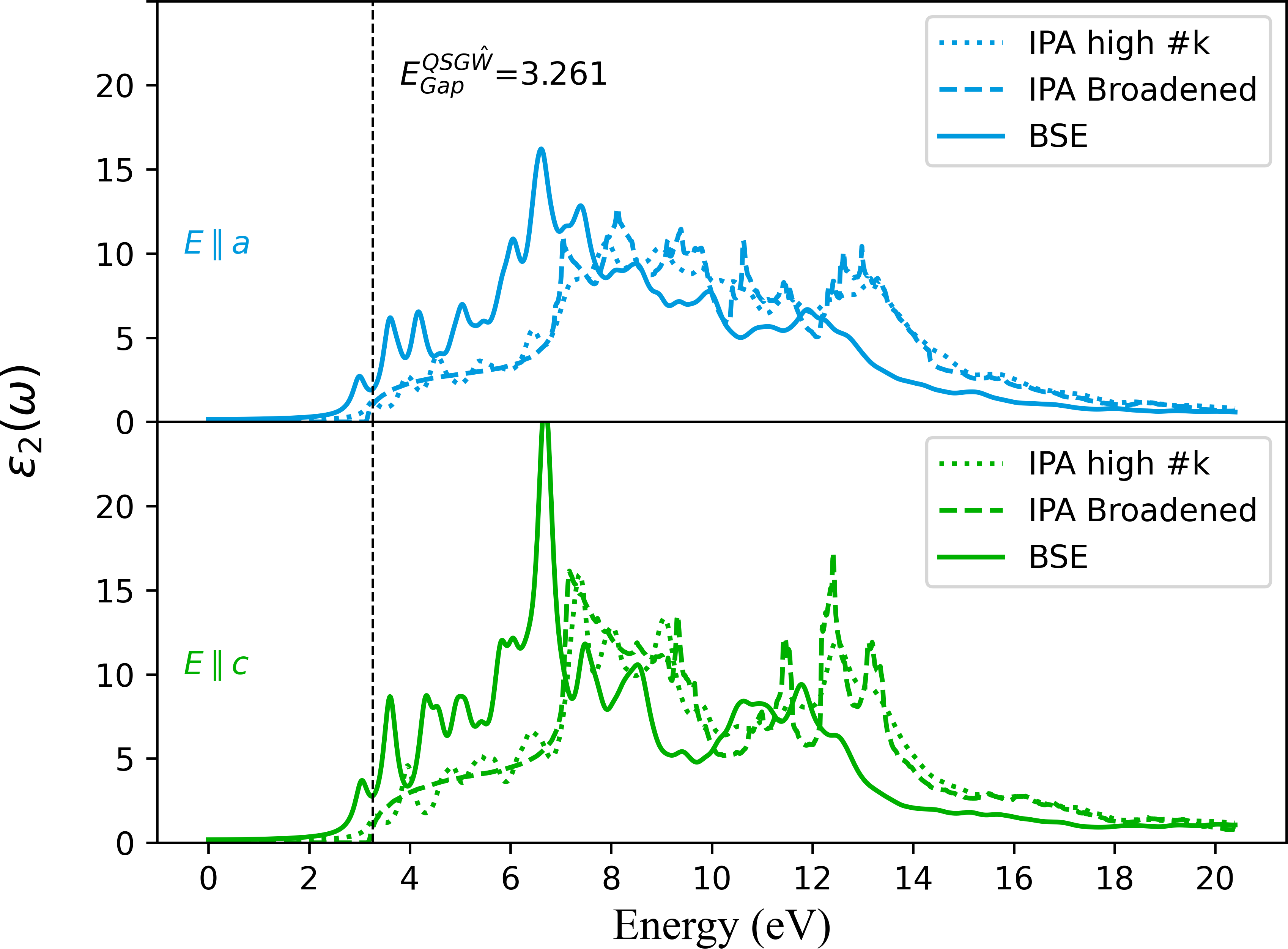}
        \caption{Imaginary part of the dielectric function in BSE and IPA for GaN for
                 polarizations perpendicular and parallel to the {\bf c}-axis. The
                 imaginary part of the energy was set to 0.01 Ry. The IPA is once
                 calculated via Eq.~\ref{eq:AdlerWiser} with a $32\times32\times16$
                 {\bf k}-mesh and the tetrahedron integration method, and also with an
                 imaginary part of $\omega$ and a $12\times12\times6$ mesh. The latter
                 {\bf k}-mesh was also used in BSE calculation.\label{figepsgan}}
    \end{figure}

  A similar overall behavior of the dielectric function occurs in GaN, as seen in
  Fig.~\ref{figepsgan}. We thus expect that there is a similar underlying physics going on
  in the band-to-band contributions and the BSE effects. We now discuss the origin of
  these observations. Because of the smaller number of bands, the analysis in terms of
  separate band-to-band transitions and where they mostly occur from in {\bf k}-space is
  somewhat easier to analyze in GaN. We will then use the band folding of the wurtzite to
  $Pna2_1$ Brillouin zone to translate these findings to ZnGeN$_2$.

\begin{figure}
      \includegraphics[width=8.5cm]{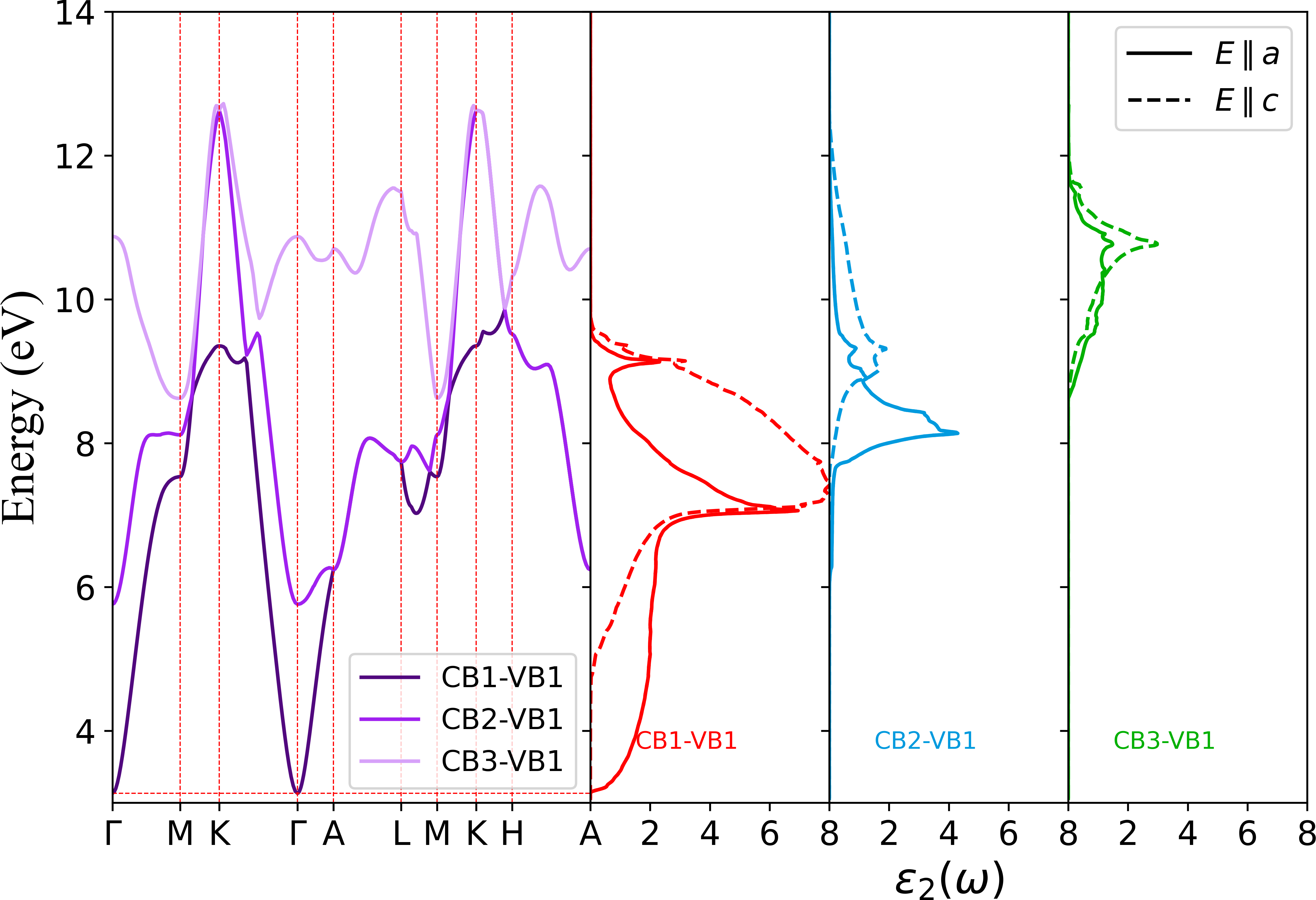}
        \caption{Imaginary part of the dielectric function $\varepsilon_2(\omega)$ of GaN
                 in the IPA associated with the transitions from the VBM and the lowest
                 three conduction bands (right), and the differences between the band
                 pairs along symmetry lines of the Brillouin zone. Higher $\varepsilon_2$
                 values expected where the band differences (solid purple lines in the
                 left panel) are flat. \label{b2bgan}}
    \end{figure}

  In Fig.~\ref{b2bgan} we show the contribution to the dielectric function in the IPA from
  the top valence band (VB$_1$) to conduction bands 1-3 (CB$_{1-3}$) along with the
  vertical difference in the bands at each {\bf k}-point along the symmetry lines. We can
  see that the onset comes only from VB$_1\rightarrow$CB$_1$ transitions and is mainly
  polarized ${\bf E}\perp{\bf c}$. This is consistent with a strongly dispersing lowest
  conduction band near $\Gamma$ and a gradual $\sqrt{\omega}$ increase of
  $\varepsilon_2(\omega)$ in the IPA. The IPA $\varepsilon_2(\omega)$ is essentially a
  matrix element modulated JDOS which is proportional to
  $\sum_{vc{\bf k}} \delta(\omega-(\epsilon_{c{\bf k}}-\epsilon_{v{\bf k}})))$. One
  expects the JDOS contributions from different band pairs $vc$ to show peaks when this
  energy band difference is nearly parallel over some region of {\bf k}-space. However,
  the matrix elements play a significant role in decreasing the intensity as energy
  increases and the $\varepsilon_2(\omega)$ shows a clearer peak structure than the JDOS.
  We thus focus directly on the IPA $\varepsilon_2(\omega)$ rather than the JDOS. The JDOS
  is shown and further discussed in Supplemental Material \cite{supmat}.
 
  Returning to Fig.~\ref{b2bgan}, the peak starting at $\sim$7.0 eV is seen to be related
  to a relatively flat energy band difference along the $M-L$ line and the onset of that
  peak coincides with the minimum of the CB$_1-$VB$_1$ energy difference along $M-L$,
  which is a saddle point in the interband energy difference surface. We can thus conclude
  that the electron-hole interaction effects in BSE shift the oscillator strength toward
  the van Hove singularity saddle point. This provides an {\sl a-posteriori} justification
  for the common practice of analyzing the optical dielectric function in terms of a
  critical point analysis, emphasizing the behavior near high symmetry points where
  Hove-type singularities occur \cite{Cardona69,Aspnes67,Aspnes80}. This was previously
  discussed in Ref. \onlinecite{Lambrecht2002} and is here confirmed. It was noted in
  \cite{Lambrecht2002} that the electron and hole have exactly the same velocity at these
  points, which could heuristically explain why they might plausibly interact stronger and
  make a large contribution when electron-hole pair interactions are included. A full
  analytical explanation of this observation would require an analysis of the matrix
  elements of $W$ as function of {\bf k} near these singularities, which is a challenging
  task and has not yet been done. However, one does expect that $W$ shifts the transitions
  down to lower energy, which explains why the peak in IPA is red-shifted in the BSE.
  Various interband transitions close in energy to each other in this region of high IPA
  contribute to a ``continuum'' exciton like peak in the BSE just below or at the saddle
  point. In other words, the two particle Hamiltonian or ``exciton'' Hamiltonian has
  eigenvalues more concentrated in energy and right at the singularity that derive from a
  wider region of {\bf k}-space individual band to-band transitions. 

  The CB$_2$ is close to CB$_1$ along $M-L$ and also contributes to the peak within the
  range 7-9 eV. There is also a rather flat interband difference of CB$_2$-VB$_1$ along
  $\Gamma-M$ near $M$. This appears at about 8 eV and contributes to the second peak in
  the IPA dielectric function. Its relation to the BSE peaks is not entirely clear because
  in the BSE formalism one can no longer associate peaks to particular band-to-band
  transitions. On the other hand, CB$_3$ is related to flat band regions along
  $\Gamma-A-L$ and contributes to the peaks near 11 eV.

\begin{figure}
      \includegraphics[width=8.5cm]{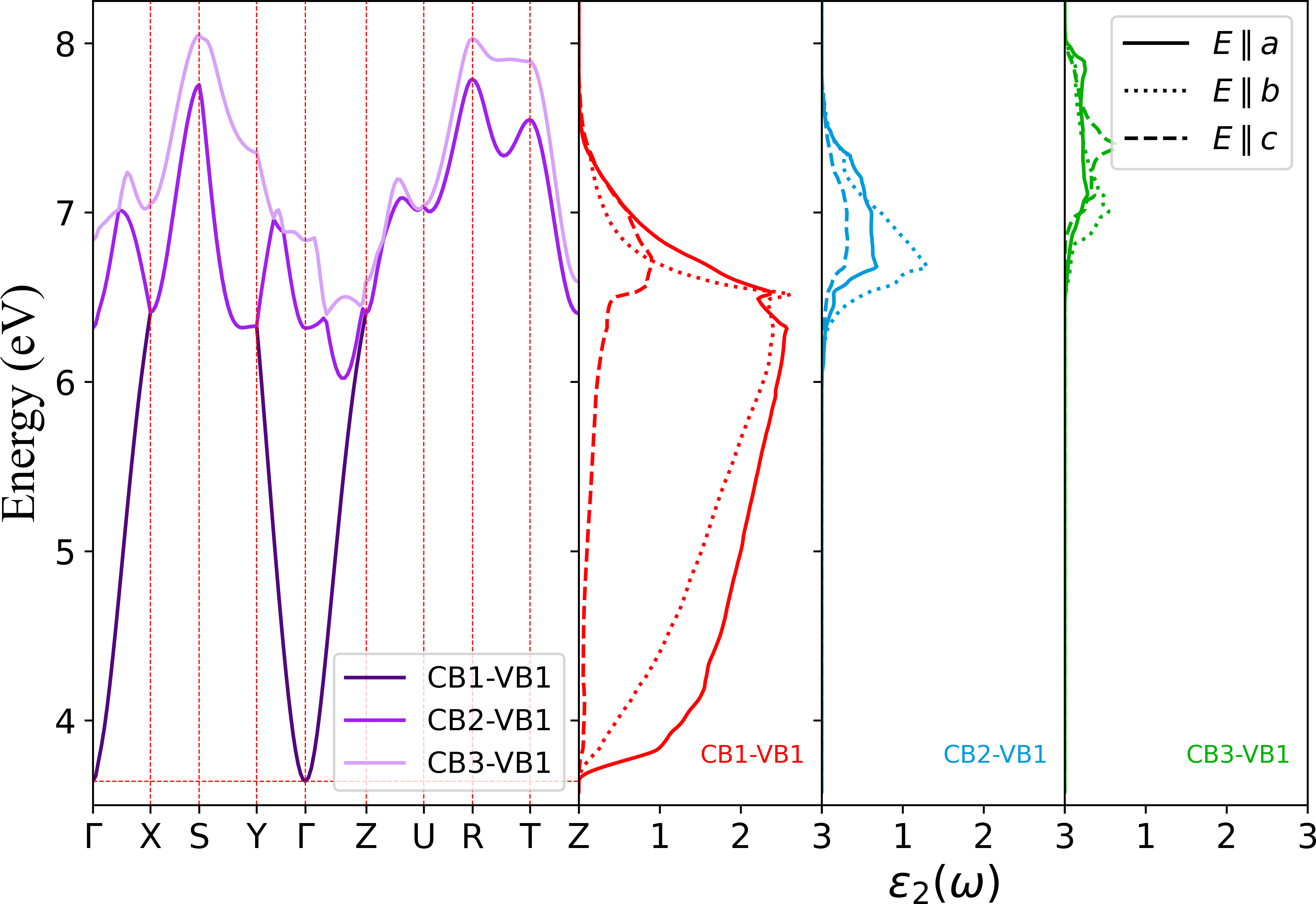}
        \caption{Imaginary part of the dielectric function $\varepsilon_2(\omega)$ of
                 ZnGeN$_2$ in the IPA associated with the transitions from the VBM and the
                 lowest three conduction bands (right) and the differences between the
                 band pairs along symmetry lines of the Brillouin zone. Higher
                 $\varepsilon_2$ values expected where the band differences (solid purple
                 lines in the left panel) are flat. \label{b2bzgn}}
    \end{figure}

  A similar behavior is seen in ZnGeN$_2$. Fig.~\ref{b2bzgn} shows various contributions
  VB$_1$-CB$_n$ to $\varepsilon_2$ for three polarizations. Additional figures are shown
  for the next top valence bands in Supplemental Material \cite{supmat}. We can again
  first see a relatively slow onset from the strongly dispersing conduction band near
  $\Gamma$. It is primarily polarized along $x=a$ near its onset consistent with the
  symmetry of VB$_1$ (which is $b_1$ or $x$-like) and the $s$-like ($a_1$ symmetry of
  CB$_1$ at $\Gamma$). A peak starts emerging at about 6.5 eV and has contributions from
  CB$_2$ and CB$_3$. The onset of the peak near 6 eV is similar to what we just discussed
  for GaN but now is related to van Hove singularities and nearly parallel bands along
  $\Gamma-Z$ which is the folded version of $ML$ in the wurtzite. One may also note at
  rather flat energy band difference of CB$_1$-VB$_1$ along $S-Y$ near $Y$. This could be
  related to the discontinuity in slope for the corresponding contribution to the IPA
  $\varepsilon_2(\omega)$ for ${\bf E}\parallel{\bf a}$ (red solid line). However, it has
  only minor effects on the BSE dielectric function. 

  In summary of this section, the major peak structure of the $\varepsilon_2(\omega)$ in
  ZnGeN$_2$ has similar origins to that in GaN and can be explained in terms of the
  band-folding effects. The peaks in the IPA correspond to regions of high JDOS (modified
  by matrix elements which decrease with increasing energy) where various bands are close
  to being parallel and allowed dipole matrix elements. The electron-hole effects further
  shift the oscillator strength towards the van Hove singularities, where the bands become
  exactly parallel, which facilitates the stronger electron-hole coupling.
  
\subsection{Static dielectric constant $\varepsilon_1(\omega=0)$} \label{sec:staticeps}
  We also calculate the real part of the dielectric function
  $\varepsilon_1 (q\rightarrow 0,\omega=0)$. In principle, we can obtain it directly from
  a Kramers-Kronig transformation of the results of Fig.~\ref{figepsbse}. However, this
  quantity is quite sensitive to the accuracy of the matrix elements of the velocity
  operator because through the Kramers-Kronig relation,
  \begin{equation}
    \varepsilon_1(0)={\cal P}\int_0^\infty \frac{2\omega d\omega}{\pi} \frac{\varepsilon_2(\omega)}{\omega^2}
  \end{equation}
  it depends on an integral over $\varepsilon_2(\omega)$ for all $\omega>0$ and is thus
  sensitive to the values of $\varepsilon_2$ not just where its peaks occur. Because of
  the difficulties in incorporating the self-energy contribution to the velocity matrix
  elements, it is more accurate to directly examine $\varepsilon_1(q,0)$ for finite $q$
  values and then extrapolate numerically to zero. Furthermore, we obtain additional
  information on the ${\bf q}$ dependence by looking at $q$ throughout the Brillouin zone.
  
\begin{figure}
      \includegraphics[width=8cm]{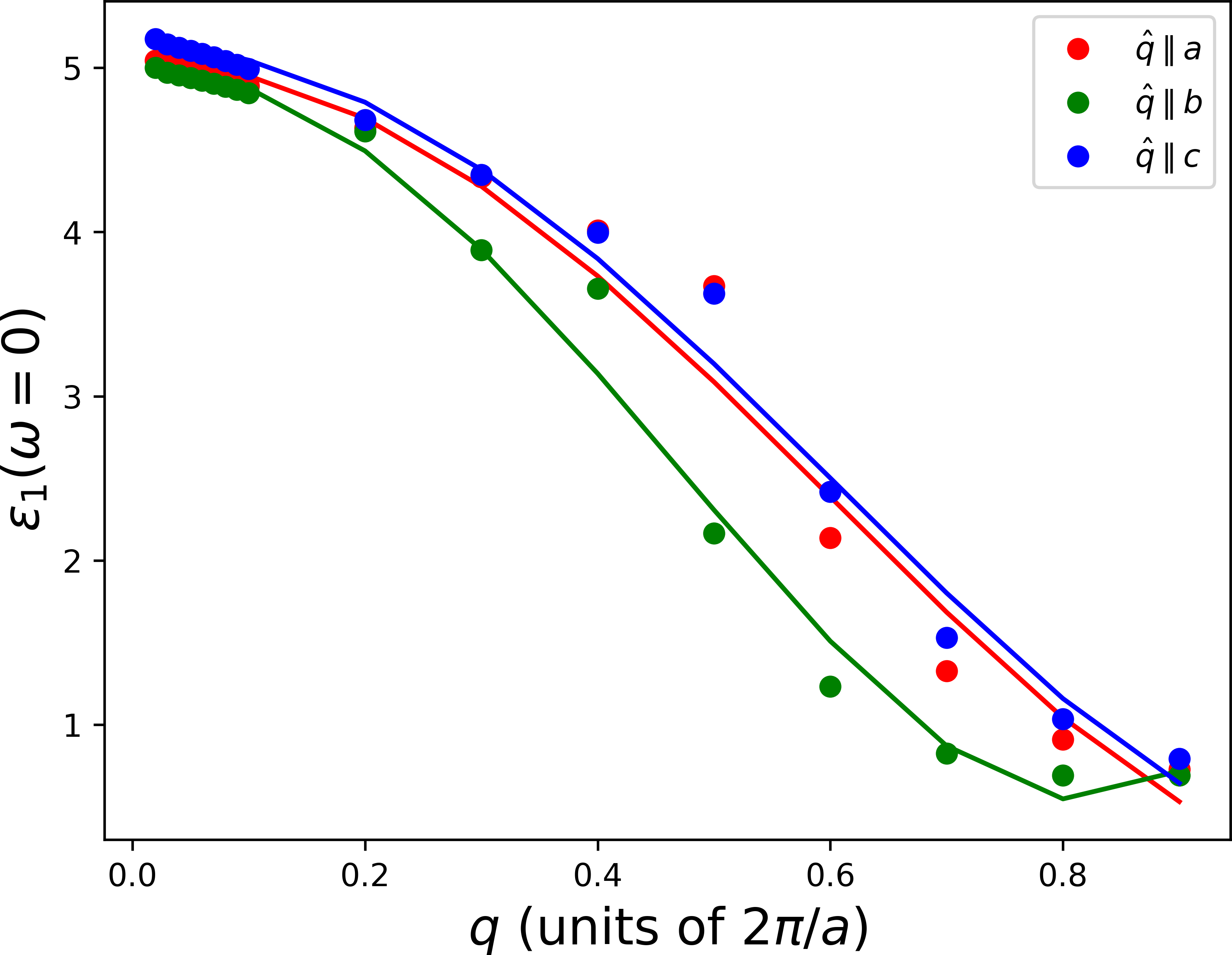}
        \caption{$\varepsilon_1(\omega = 0)$ of ZnGeN$_2$ is plotted with respect to
                 $q$-points around $q=0$.\label{fig:eps1}}
    \end{figure}

  We use a model dielectric function which was introduced by Cappellini \etal in
  \cite{Cappellini93}, to fit to the directly calculated results from the BSE at finite
  {\bf q} shown as filled circles as shown in Fig.~\ref{fig:eps1}. Extrapolation of this
  fit to $q=0$ limit gives the $\varepsilon_1 (q=0,\omega=0)$ values of 5.04 along $x$,
  5.01 along $y$, and 5.13 along $z$ directions. These values of course correspond to
  electronic screening only, and therefore, to what is commonly referred to as
  $\varepsilon^\infty$, rather than the true static value which would include the phonon
  contributions. In other words, it applies to the frequency range well below the band gap
  but high above the phonon frequencies and is related to the index of refraction measured
  in this region by $n=\sqrt{\varepsilon_1}$. The indices of refraction are thus 2.24,
  2.24, 2.26 for $x,y,z$ directions. Our values of
  $\varepsilon_1({\bf q}\rightarrow0,\omega=0)$ obtained here are close to those obtained
  from a Berry phase calculation of the polarization in the LDA \cite{Paudel08} which are
  5.232, 5.166, 5.725 for $a,b,c$ directions respectively, or corresponding to indices of
  refraction of 2.29, 2.26 and 2.39. On the other hand, if we had used the
  $\varepsilon_1({\bf q}=0,\omega=0)$ obtained directly from the Kramers-Kronig
  transformation of the BSE results in Fig.~\ref{figepsbse}, we would find
  $\varepsilon^\infty$ values of 9.79, 9.02 and 9.30 for the $a$, $b$, $c$ directions.
  These are significantly overestimated by almost a factor 2 and this results from the
  incorrect velocity matrix elements when taking the limit ${\bf q}\rightarrow0$
  analytically. Furthermore, the values obtained by the numerical extrapolation are close
  to those for GaN which is about 5.35 and with only small anisotropy for the high
  frequency dielectric constant. They are slightly smaller than the values obtained by
  calculating the adiabatic response to a static electric field using Berry phase
  calculations of the polarization as reported in \cite{Paudel08}, which used the local
  density approximation.    

\subsection{Valence band splittings and excitons} \label{sec:split}
    \begin{figure*}
      \includegraphics[width=5cm]{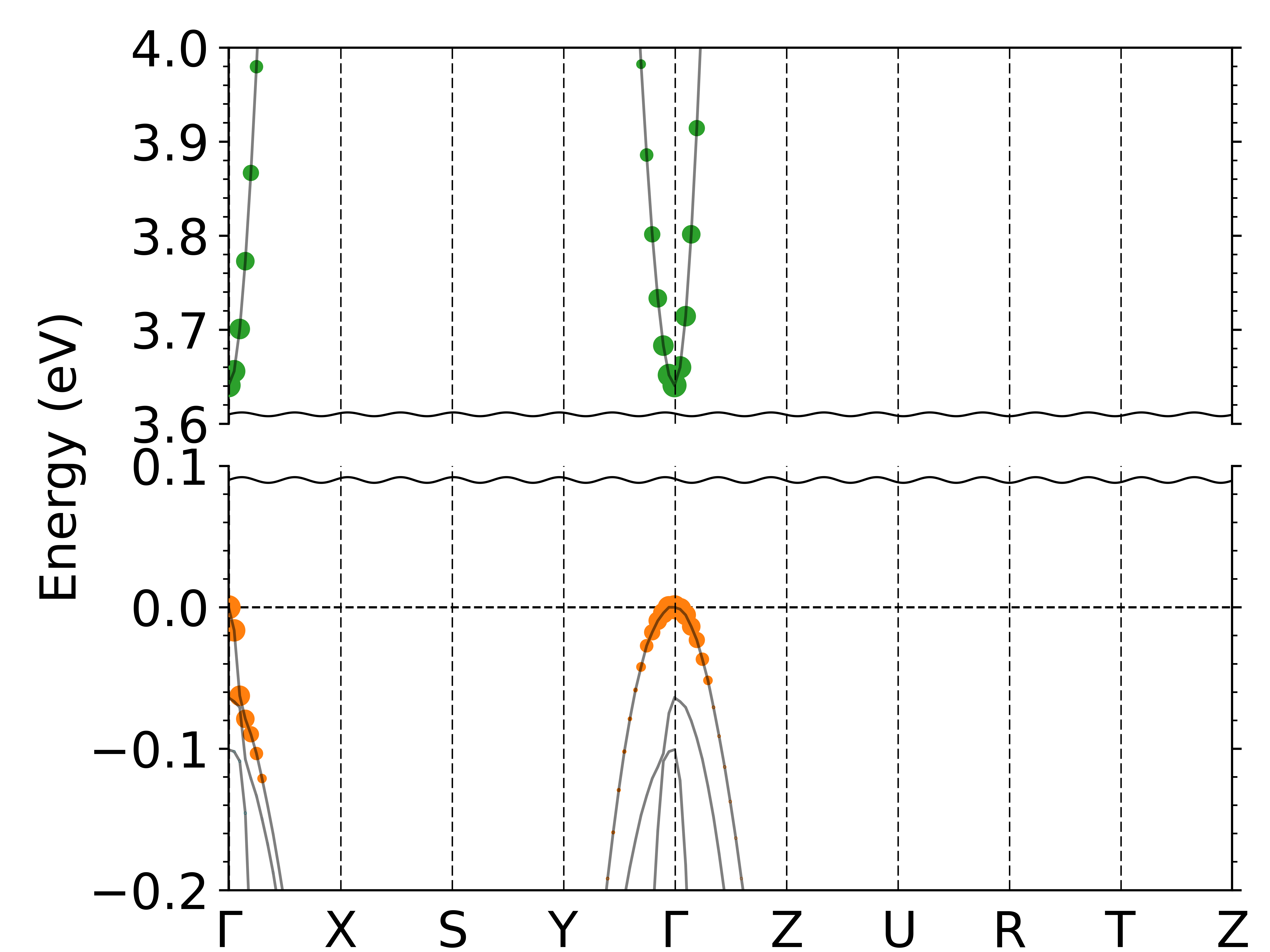}
      \includegraphics[width=5cm]{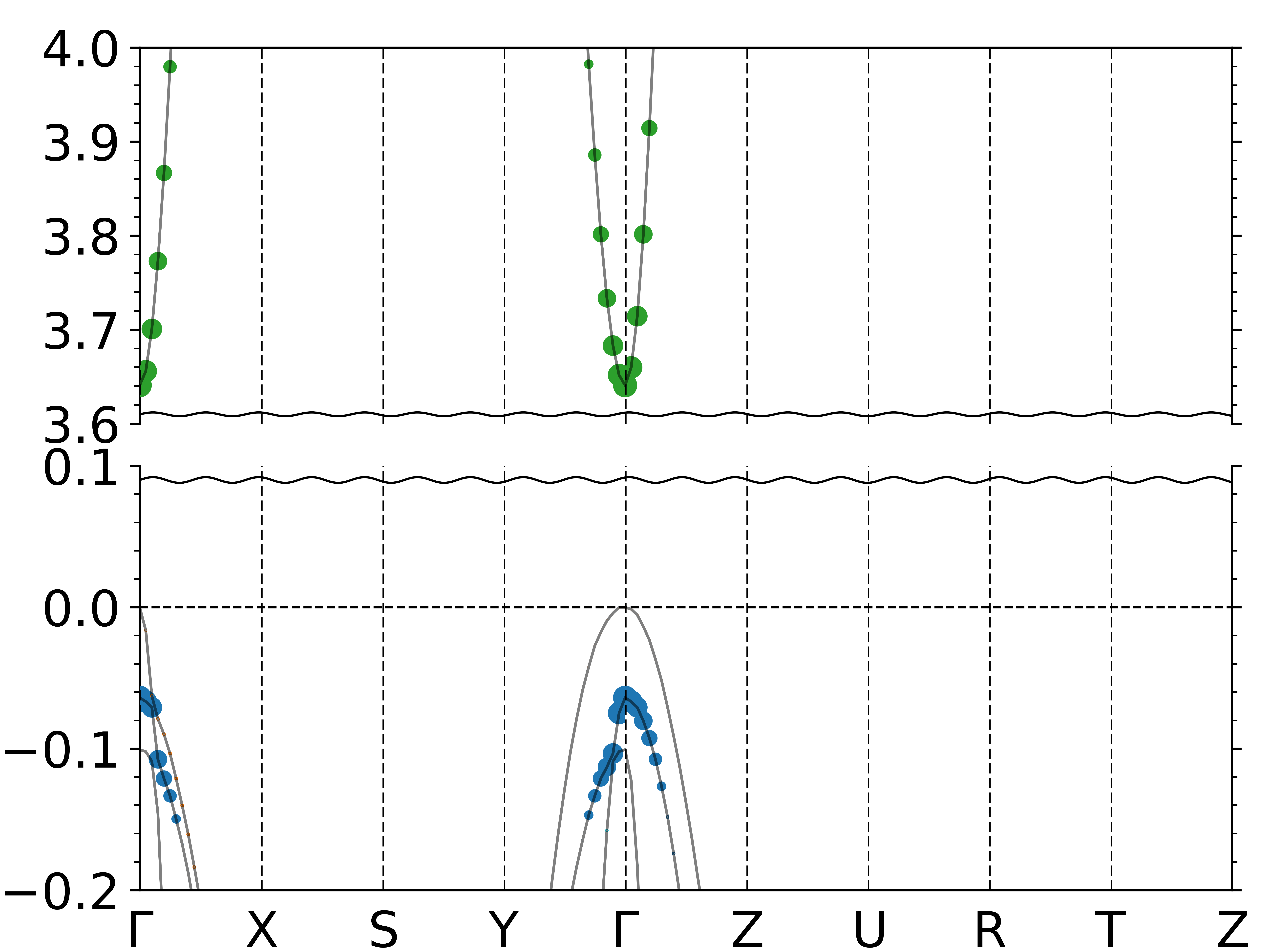}
      \includegraphics[width=5cm]{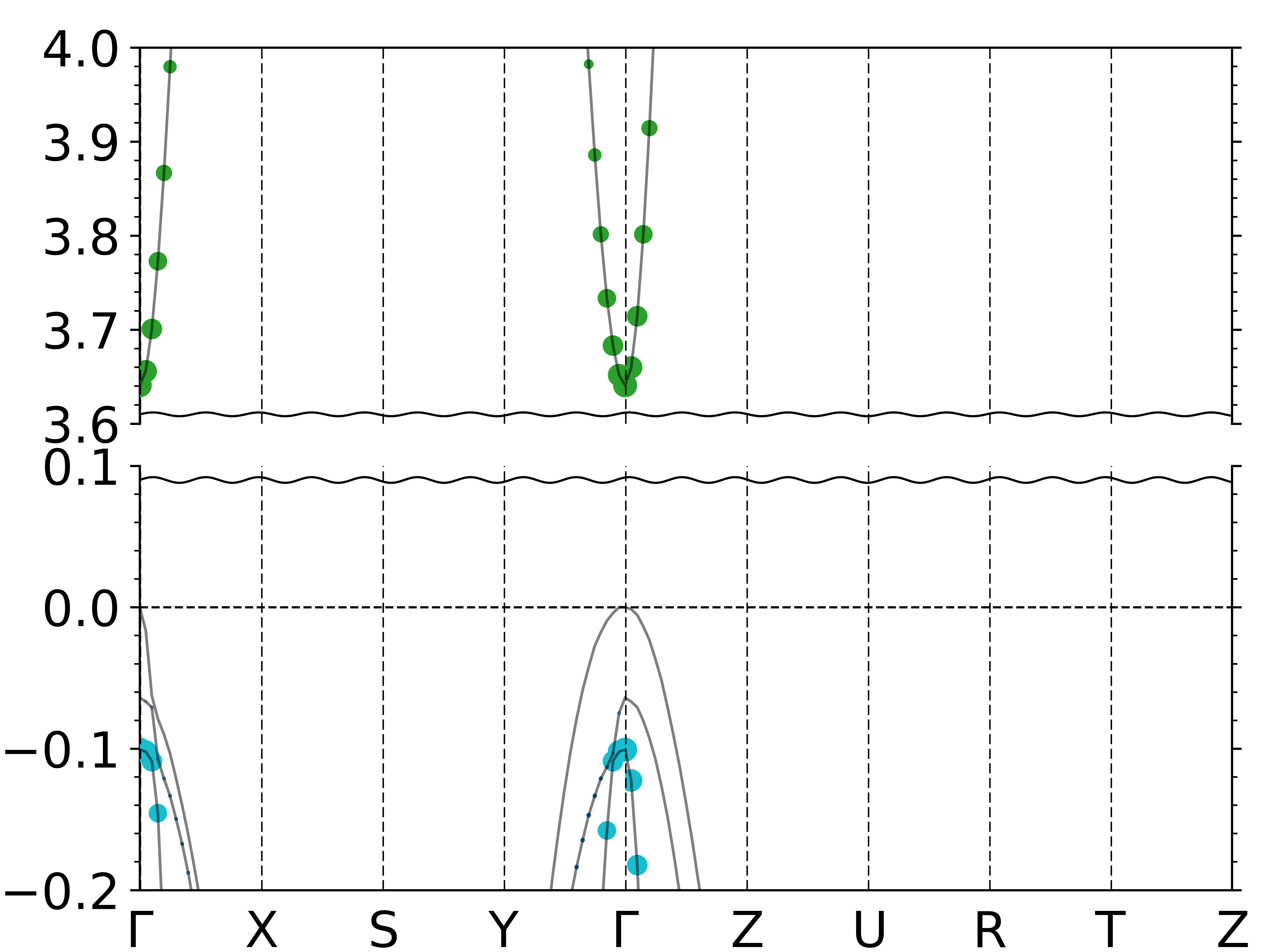}
        \caption{Exciton band weights for the lowest three excitons, associated each with
                 a single valence band. Note that $N_v=24$ and $N_c$=16 bands were included
                 but only the bands shown contribute significantly to these excitons.
                 \label{figbndexg}}
    \end{figure*}
  In this section, we turn our attention to the energy region below the band-to-band
  continuum. Using a smaller broadening or examining the exciton eigenvalues below the
  quasiparticle gap directly, we find that there are three bright excitons, each polarized
  along a different direction, from lower to higher energy $x$, $y$ and $z$. The excitons
  for each polarization clearly correspond to transitions from each of the top three
  valence bands to the CBM and are consistent with the symmetry analysis of these bands
  given below. The exciton energies, however, are found to be sensitive to the
  {\bf k}-mesh used in the BSE equation. For the BSE dielectric function calculation
  over the full $\omega$-range as shown in Fig.~\ref{figepsbse} we used a
  $N_k\times N_k\times N_k$ mesh with $N_k=6$ and with $N_v=24$ and $N_c=16$.
  Subsequently, we used different meshes with $N_k$ varying from 3 to 6 to study the
  exciton convergence while maintaining the same $N_v,N_c$. Separately, because we found
  (see Fig.~\ref{figbndexg}) that these excitons arise primarily from the top three
  valence bands and the lowest conduction band, we solved the BSE equation for only
  $N_v=3$ and $N_c=1$, but now going up to a $N_k=10$ mesh. The extrapolation to
  $N_k\rightarrow\infty$ can be done in various ways, either fitting to an exponential
  saturation curve or by plotting versus $1/N_k$ and extrapolating the points with
  smallest $1/N_k$ to zero with a polynomial. The details of our extrapolation are
  described in Supplemental Material \cite{supmat}.
 
  We obtain a lowest exciton gap of 3.51$\pm0.01$ eV. Comparing with the quasiparticle gap
  of 3.64 eV, this gives an exciton binding energy of 0.13$\pm0.01$ eV. To the precision
  of the calculations, the same exciton binding energy is obtained for the excitons
  polarized in the other directions. These exciton binding energies are overestimated
  because they only include electronic screening. Within a hydrogenic model of the
  Wannier-Mott exciton, the binding energy would be given by an effective Rydberg divided
  by the dielectric constant squared and multiplied by a reduced effective mass. We can
  thus correct for our use of only electronic screening by multiplying by a scaling factor
  $\varepsilon^2_{\infty}$/$\varepsilon^2_0$. Table~\ref{table:scaling_BE} shows the
  $\varepsilon_{\infty}$ and $\varepsilon_0$ values from the literature and our scaled
  binding energies. The high frequency dielectric tensor of GaN was calculated by various
  authors using density functional perturbation theory \cite{Karch98,Bungaro2000}. The
  phonon calculations in these papers provide $\omega_{LO}/\omega_{TO}$ for
  ${\bf E}\parallel {\bf c}$ from the $A_1$ modes and for ${\bf E}\perp{\bf c}$ from the
  $E_1$ modes and through the Lyddane-Sachs-Teller relation,
  $\varepsilon_0/\varepsilon_\infty=(\omega_{LO}/\omega_{TO})^2$ provide us the scaling
  factor needed. In Table~\ref{table:scaling_BE} we used the values from Karch \etal
  \cite{Karch98} but to the precision we here require, there is good agreement also with
  the results extracted from Bungaro \etal \cite{Bungaro2000} and also with the
  experimental results $\varepsilon_\infty\approx 5.35$ for both parallel and
  perpendicular to ${\bf c}$ directions, and
  $\varepsilon_0^\perp=9.5$, $\varepsilon_0^\parallel=10.4$ given by Barker and Ilegems
  \cite{Barker73}. We can see from Table~\ref{table:scaling_BE} that the
  $\varepsilon_\infty/\varepsilon_0$ is about 0.55$\pm0.01$ and
  $[\varepsilon_\infty/\varepsilon_0]^2\approx0.3$ and hence we obtain a binding energy of
  about 0.04$\pm0.02$ eV. Similar results are obtained for GaN and the details of these
  calculations are given in the Supplementary Material \cite{supmat}. The resulting
  estimates for the exciton binding energy are of the right order of magnitude and
  consistent with the known values for GaN of 22-26 meV \cite{Reimann1998}. However, the
  accuracy with which we can obtain these exciton binding energies is still limited by the
  number of k-points we can afford, the number of bands we can include concurrently, and
  by the limitations of the present BSE implementation which does not include lattice
  dynamical effects. 

    \begin{table}[h]
      \caption{Literature values of static dielectric constant $\varepsilon_0$, high
               frequency dielectric constant $\varepsilon_{\infty}$, and the scaled
               binding energies. The estimated uncertainty results primarily from the
               uncertainties in the extrapolation.\label{table:scaling_BE}}
      \begin{ruledtabular}
        \begin{tabular}{lccccc}
                                     &                           & $\varepsilon_0$ & $\varepsilon_{\infty}$ & $(\varepsilon^2_{\infty}$/${\varepsilon^2_0})$ & BE (meV)        \\\hline
          \multirow{3}{*}{ZnGeN$_2$} & E$\parallel$a$^{\ast}$    & 9.28            & 5.24                   & 0.32                                           & 0.044$\pm$ 0.003\\
                                     & E$\parallel$b$^{\ast}$    & 9.22            & 5.17                   & 0.31                                           & 0.042$\pm$ 0.003\\
                                     & E$\parallel$c$^{\ast}$    & 10.61           & 5.73                   & 0.29                                           & 0.040$\pm$ 0.003\\\hline
          \multirow{2}{*}{GaN}       & E$\parallel$c$^{\dagger}$ & 10.3            & 5.41                   & 0.28                                           & 0.042$\pm$ 0.001\\
                                     & E$\perp$c$^{\dagger}$     & 9.22            & 5.21                   & 0.32                                           & 0.049$\pm$ 0.001\\
        \end{tabular}
          \begin{tablenotes}
            \small
              \item $^{\ast}$ Ref. \cite{Paudel08}
              \item $^{\dagger}$ Ref. \cite{Karch98}
          \end{tablenotes}
      \end{ruledtabular}
    \end{table}
    
  To further prove that these excitons are associated with each of the valence bands, the
  contribution of the valence and conduction bands in a band plot to each exciton are
  shown in Fig.~\ref{figbndexg}. Again, this is obviously consistent with their
  polarization and the symmetry of the valence bands at $\Gamma$.

\begin{figure*}
      \includegraphics[width=8cm]{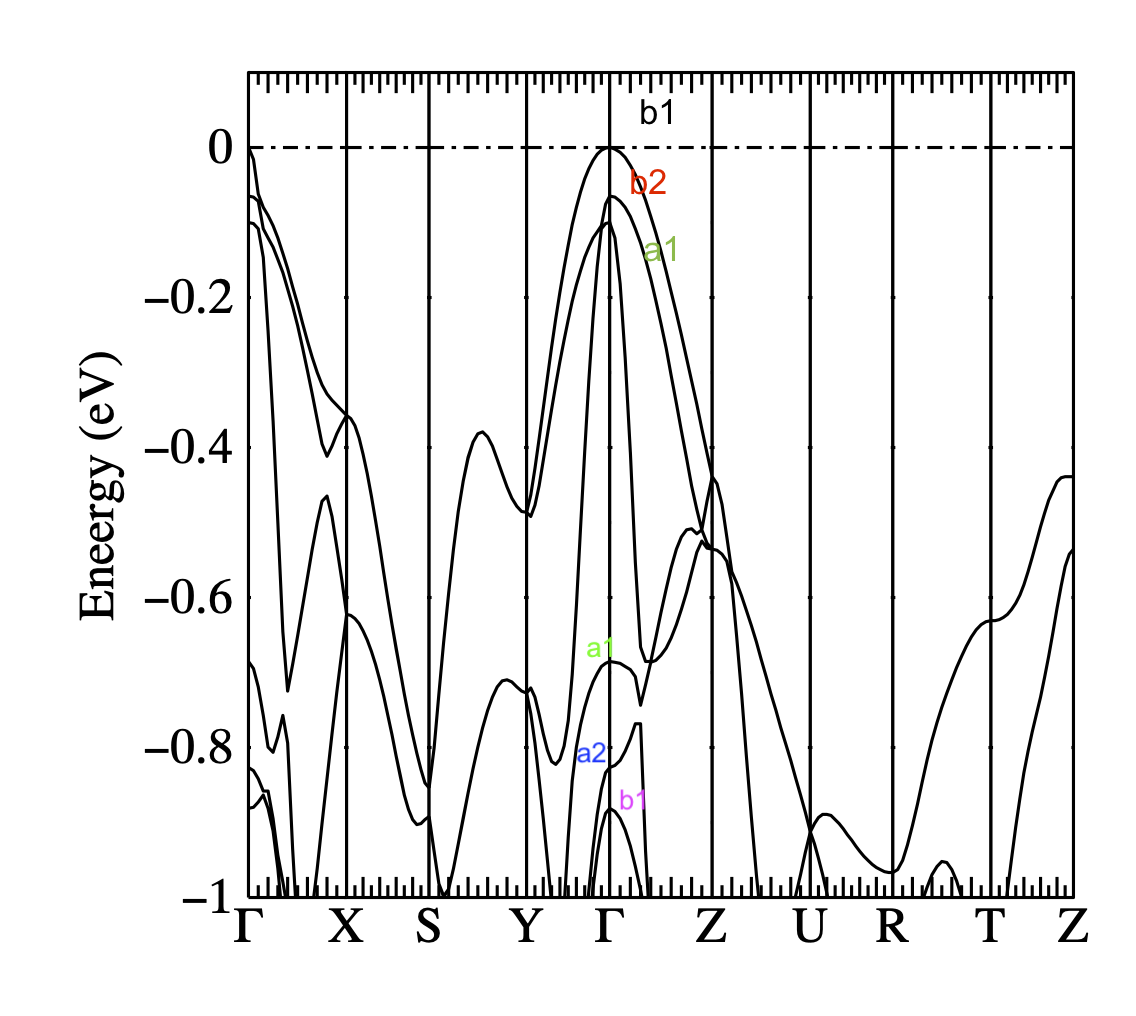}
      \includegraphics[width=7.5cm]{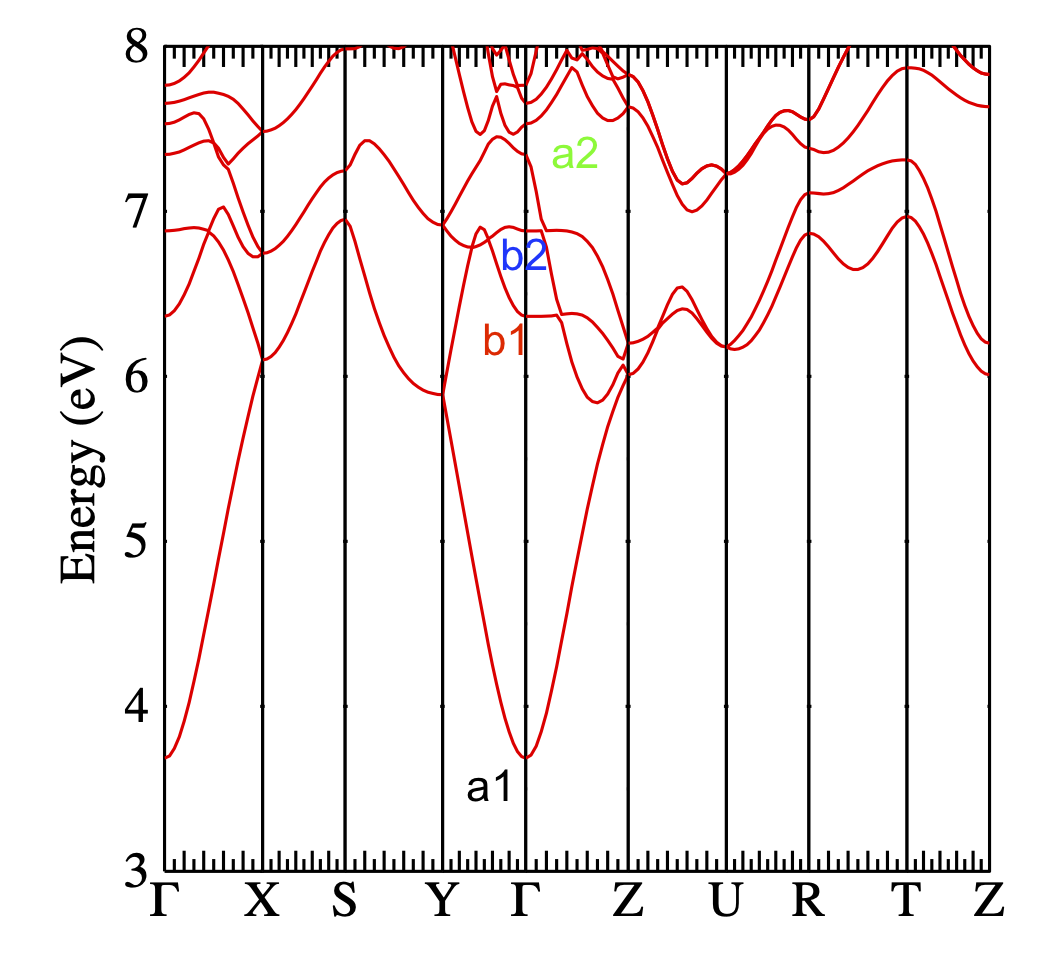}
        \caption{Zoom in on the valence band maximum region (left) and conduction band
                 range (right) of ZnGeN$_2$ as obtained in QS$G \hat W$. The bands at
                 $\Gamma$ are labeled by their irreducible representations.
                 \label{figbndzoom}}
    \end{figure*}

  We now return to the valence band splitting itself. In Fig.~\ref{figbndzoom} we show a
  zoom in on the valence band maximum energy range as well as the conduction band region.
  The bands are labeled at $\Gamma$ by their irreducible representations in the point
  group $C_{2v}$. The VBM has $b_1$ symmetry, which corresponds to $x$, and transitions
  from this band to the CBM, which has $s$-like $a_1$ ($\Gamma_4$ in Koster\etal notation
  \cite{Koster}) symmetry, are thus allowed for ${\bf E}\parallel{\bf a}$. The next band
  has symmetry $b_2$ ($\Gamma_2$) corresponding to $y$ or ${\bf b}$ and the third band has
  $a_1$ ($\Gamma_1$) symmetry corresponding to $z$ or ${\bf c}$. Note that although this
  ordering appears to be the same as in Ref.\onlinecite{Punya11}, it is in fact different
  because a different setting of the space group, namely $Pbn2_1$ was used in that paper,
  which means that the $a$ and $b$ directions are reversed from the present paper and
  hence also $x$ and $y$ or $b_1$ and $b_2$. This difference is due to a difference in the
  lattice constant ratio $a/b$ and already indicates that this splitting is sensitive to
  strain. We will discuss this strain dependence in the following section. The next few
  valence bands are also labeled in Fig.~\ref{figbndzoom}. The splittings between the top
  VBM and the next two bands are given in Table~\ref{tabsvsplit}. The next higher
  conduction bands have $b_1$, $b_2$ and $a_2$ symmetry and are situated at 2.68, 3.20 and
  3.66 eV higher in energy, respectively. 

    \begin{table}[h]
      \caption{Valence band splittings in meV in ZnGeN$_2$.\label{tabsvsplit}}
      \begin{ruledtabular}
        \begin{tabular}{lccc}
                    & GGA & QS$GW$ & QS$G\hat W$ \\ \hline
          $b_1-b_2$ & 72  & 64     & 65          \\
          $b_1-a_1$ & 94  & 102    & 101 
        \end{tabular}
      \end{ruledtabular}
    \end{table}

  In terms of the valence bands, one may notice that the top valence band with $b_1$ or
  $x$-like symmetry has the smallest hole mass in the $x$-direction, while the next one
  with $b_2$ or $y$-like symmetry has the lowest mass in the $y$-direction and the third
  valence band (counted from the top down) has the lowest mass in the $z$ direction. Thus,
  the symmetry of the band corresponds to the direction in which the hole mass is the
  smallest. This was also noted in a recent paper on MgSiN$_2$-GaN alloys \cite{Dernek22},
  although the explanation there is not quite correct. It can be explained in terms of the
  well-known ${\bf k}\cdot{\bf p}$ expression for the effective mass
    \begin{equation}
      M^{-1}_{\alpha\beta}= \delta_{\alpha\beta}\frac{1}{m_e} +\frac{1}{m_e^2}\sum_{n^\prime\ne n}
      \frac{\langle n{\bf k}|p_\alpha|n^\prime{\bf k}\rangle\langle n^\prime{\bf k}|p_\beta|n{\bf k}\rangle +c.c.}{E_{n{\bf k}}-E_{n^\prime{\bf k}}}
    \end{equation}
  and the matrix elements between states that are non-zero can be found by symmetry. For
  example, for the VBM of $b_1$ symmetry and focusing first on the interaction between VBM
  and CBM, momentum matrix elements are only allowed in the $x$ direction, so these bands
  are pushed away from each other along $\Gamma-X$, which reduces the hole mass in this
  direction. In the $y$ direction, only interactions with the higher $a_2$ conduction
  band; or in the $z$ direction, only interactions with the higher $b_1$ conduction band
  would come in. These will be smaller because of the larger energy denominator, hence the
  hole mass is expected to be smallest in the $x$ direction due to the strongest
  interaction with the conduction band. On the other hand, the interactions between nearby
  valence bands below the $b_1$ state would have the opposite effect of increasing the VBM
  hole mass. While the VBM splittings are much smaller, the momentum matrix elements must
  be smaller so these have less of an effect. This is because both valence bands are
  N$_{2p}$ like and intra-atomic matrix elements of the momentum operator are forbidden
  for the same angular momentum $\ell$. 

  Likewise, for the second valence band the interaction with the lowest conduction band is
  only allowed for the $y$ direction, and for the third valence band of $a_1$ symmetry
  this interaction is only allowed for the $z$-direction. On the other hand, from the CBM
  point of view, there are matrix elements with either the valence bands of symmetries
  $b_1$ in $x$-direction, with $b_2$ along the $y$ direction and with $a_1$ along the
  $z$-direction. Hence, the CBM has a nearly isotropic effective mass tensor. These
  considerations apply quite generally to the II-IV-N$_2$ semiconductors.

\subsection{Strain effects on valence band splittings.}\label{sec:strain}
\begin{figure}
      \includegraphics[width=8cm]{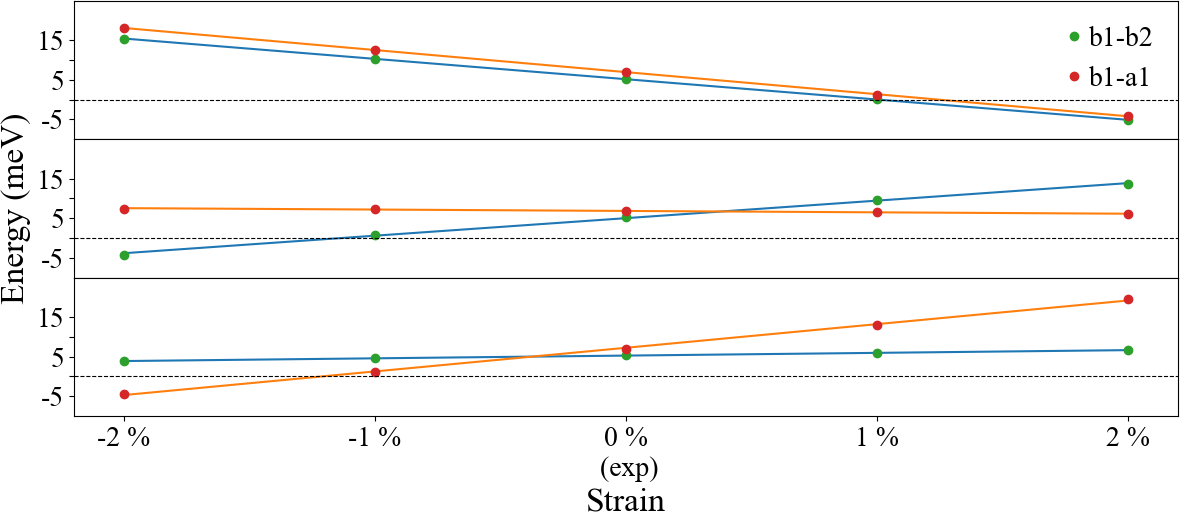}
      \caption{Splitting between the top valence bands due to strain in x-, y- and
               z-directions (from top to bottom) are observed. The percentages on the
               x-axis of the figure indicate the amount of strain applied to one major
               axis in comparison to the experimental lattice constants, while the volume
               of the unit cell kept fixed. \label{strain}}
    \end{figure}

  In Fig.~\ref{strain}, we show the strain affect on the band splittings for the top three
  valence bands. We analyze the changes in the splitting due to a uniaxial strain within
  $\pm$ 2 $\%$, applied on one major axis at a time, where negative strain means that the
  experimental lattice constants are compressed, and positive strain means they are
  stretched. The unit cell volume is kept constant at the experimental value. In other
  words, we consider pure shear strains. As shown in Fig.~\ref{figbndzoom}, the top three
  valence bands have the symmetry of $b_1$, $b_2$ and $a_1$. Fig.~\ref{strain} shows the
  splittings $b_1-b_2$ and $b_1-a_1$, following the same symmetry labeling. Based on
  symmetry grounds, we expect that with a strain along $x$, which has $b_1$ symmetry, only
  the $b_1$ eigenvalue will shift. Hence, both band splittings will shift parallel to each
  other. We can see that compressive strain along $x$ leads to a larger $b_1-a_1$ and
  $b_1-b_2$ splitting. For strain along $b$ or $y$, only $b_2$ eigenvalues should be
  affected, so the $b_1-a_1$ splitting stays constant but the $b_1-b_2$ splitting behaves
  linear with strain. Again, compression raises the $b_2$ level and hence reduces the
  $b_1-b_2$ splitting. At some point it crosses through zero at which point the VBM
  becomes the $b_2$ band. Likewise, for strain along $c$, the $a_1$ level shifts and the
  $b_1-b_2$ splitting stays constant but the $b_1-a_1$ splitting shifts linear and crosses
  through zero for a critical strain. 

\subsection{Comparison to experimental results}\label{sec:expt}
\begin{figure}[h!]
      \includegraphics[width=8cm]{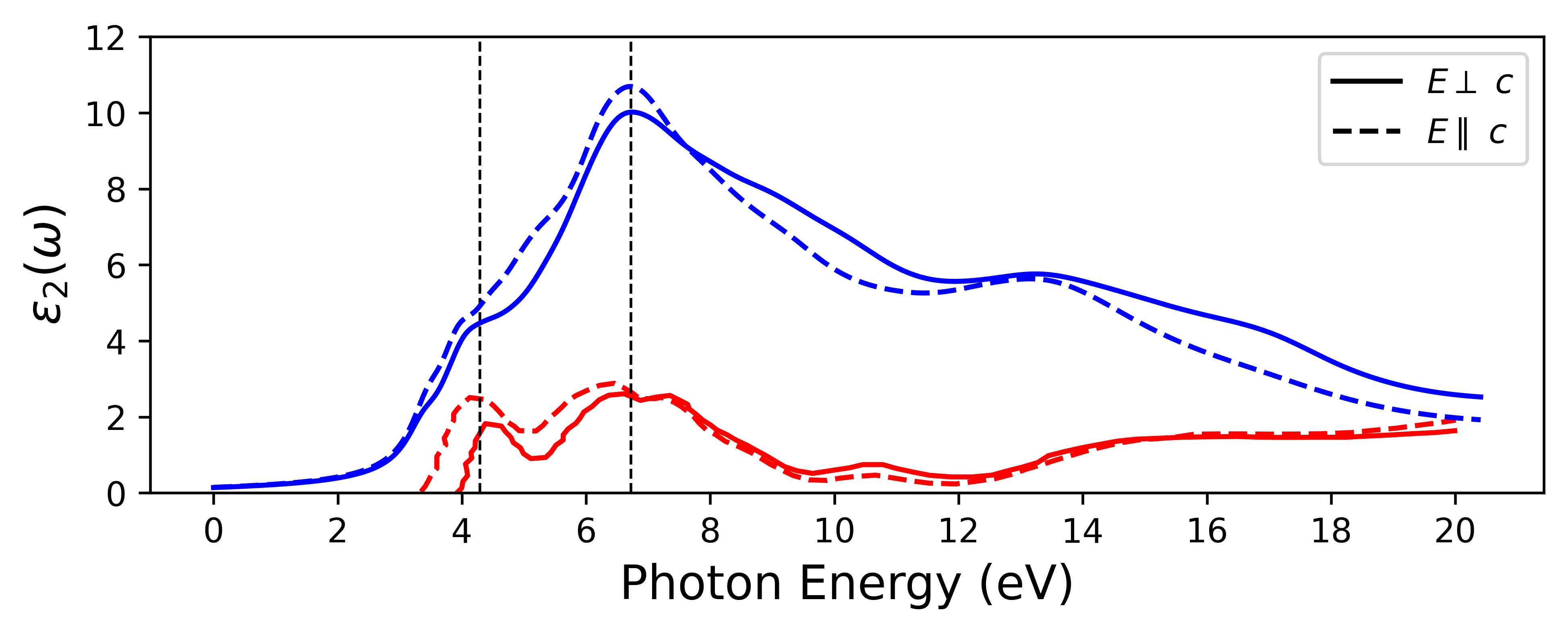}
        \caption{Comparison between calculated results (blue curves) to digitalized data
                 (red curves) from Misaki \etal \cite{Misaki03} \label{figbndexp}. We used
                 the average of $\varepsilon_2(\omega)$ in $a$ and $b$ directions shown in
                 Fig.~\ref{figepsbse}}
    \end{figure}

  We are aware of only one experimental report of the optical dielectric function in the
  visible to UV region. Misaki \etal \cite{Misaki03} reported reflectivity spectra of
  ZnGeN$_2$ parallel and perpendicular to the {\bf c} axis, then subsequently extracted
  $\varepsilon_1(\omega)$ and $\varepsilon_2(\omega)$ by using the Kramers-Kronig
  analysis. Unfortunately, the data do not distinguish the $a$ and $b$ directions in the
  $c$-plane. It is also possible that cation disorder led to effectively a wurtzite type
  structure. We therefore compare an average of $a$ and $b$ polarization with their
  results in Fig.~\ref{figbndexp}. The location of the main peaks is in reasonable
  agreement but the magnitude of $\varepsilon_2$ differs markedly between theory and
  experiment. They show a first peak near 4 eV, a main peak near 7 eV, a dip and small
  peak near 10 eV and then a plateau between 15 and 20 eV. These results are roughly
  consistent with our calculated optical properties. The experimental data show only a
  small anisotropy. The cut-off at lower energies is likely affected by the presence of
  finite film-thickness related interference effects. Our calculated values of
  $\varepsilon_2$ are notably higher than the experimental values. On the theory side,
  this could be, in part, due to an overestimate of the velocity matrix elements. This is
  known to lead to an overestimate of $\varepsilon_1(\omega=0)$ as well as we discussed
  earlier. It can be remedied by performing calculations at finite ${\bf q}$ close to
  ${\bf q}=0$ and extrapolating instead of relying on an analytical derivation of the
  limit for ${\bf q}\rightarrow0$. On the other hand, the experimental values, extracted
  from reflectivity may also be underestimated because of surface roughness, which can
  reduce the fraction of the signal that is specularly reflected \cite{Lambrecht95}.

\section{Conclusion}
  In this paper we revisited the QS$GW$ band structure calculations of ZnGeN$_2$ using the
  experimental lattice constants and an improved calculation of the screened Coulomb
  interaction which includes ladder diagrams or electron-hole effects in the polarization.
  We also calculated the optical dielectric function corresponding to interband
  transitions at both the IPA and BSE levels. The differences between the two were
  compared with those in GaN and indicate a shift of oscillator strength of the first peak
  toward the critical points near its onset. These were analyzed in terms of individual
  band-to-band transitions, taking band folding effects into account between GaN and
  ZnGeN$_2$ Brillouin zones and analysis of the IPA $\varepsilon_2(\omega)$ in terms of
  band-to-band differences in {\bf k}-space. Well defined excitons below the gap are found
  but require very fine {\bf k}k-meshes to determine with sufficient accuracy. Strain
  effects on the band splittings and hence exciton splittings resulting from the
  orthorhombic crystal field splitting were presented. 

\acknowledgments{This work was supported by the U.S. Department of Energy Basic Energy
                  Sciences (DOE-BES) under grant No. DE-SC0008933. Calculations made use
                  of the High Performance Computing Resource in the Core Facility for
                  Advanced Research Computing at Case Western Reserve University and the
                  Ohio Supercomputer Center.}

\appendix*

\bibliography{Bib/dft,Bib/gw,Bib/lmto,Bib/zgn,Bib/BSE,Bib/gan}
\end{document}

\title{Supplementary Material: Interband and exciton related optical properties of ZnGeN$_2$ compared to GaN}
\author{Ozan Dernek}
\author{Walter R. L. Lambrecht}
\affiliation{Department of Physics, Case Western Reserve University,
  10900 Euclid Avenue, Cleveland, OH 44106}
\maketitle
\section{Band folding}
    \begin{figure}
      \includegraphics[width=9cm]{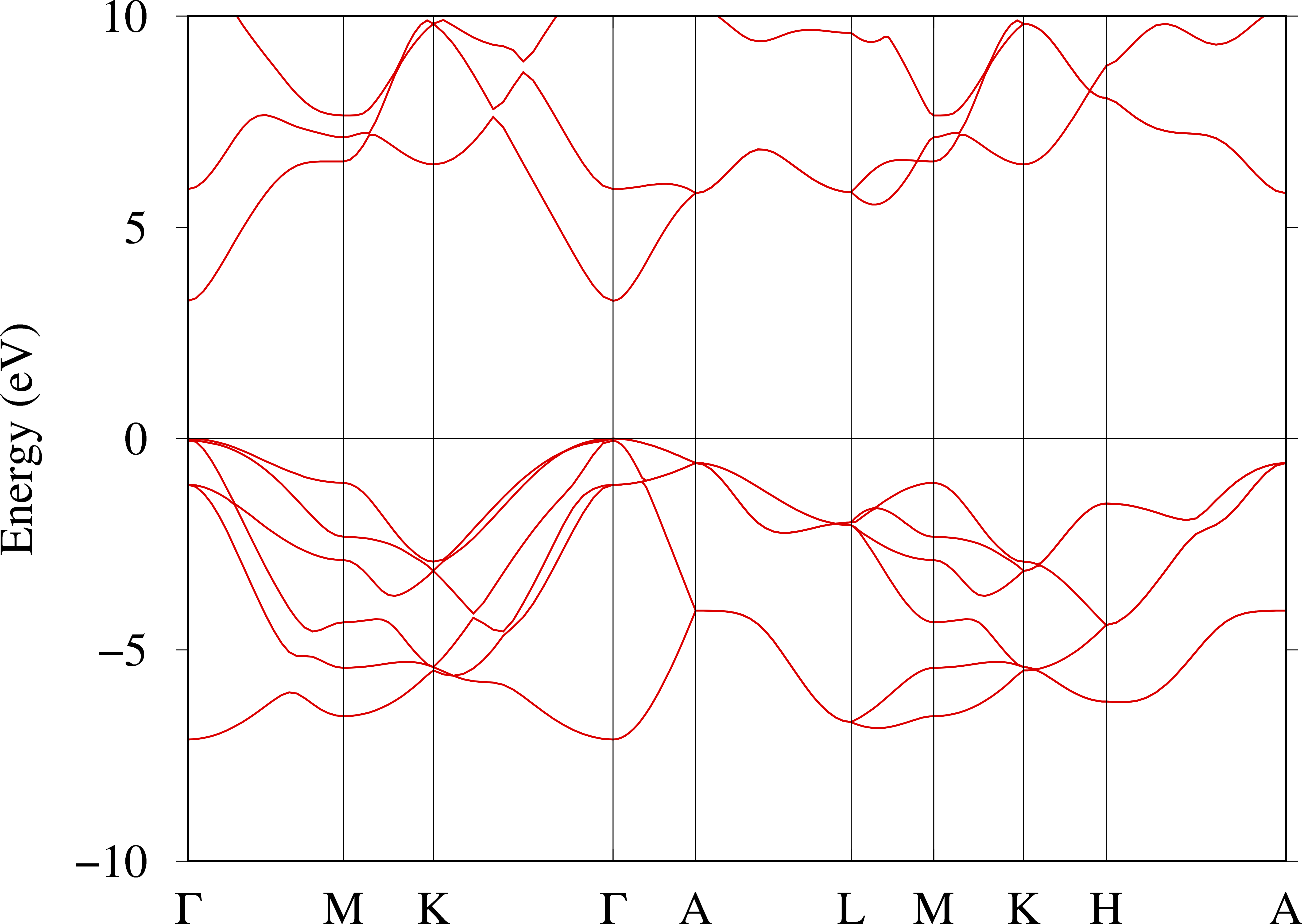}
        \caption{Band structure of GaN in QSG$\hat{W}$ approximation in the wurtzite
                 Brillouin zone.\label{figgan}}
    \end{figure}
    
\begin{figure}
      \includegraphics[width=9cm]{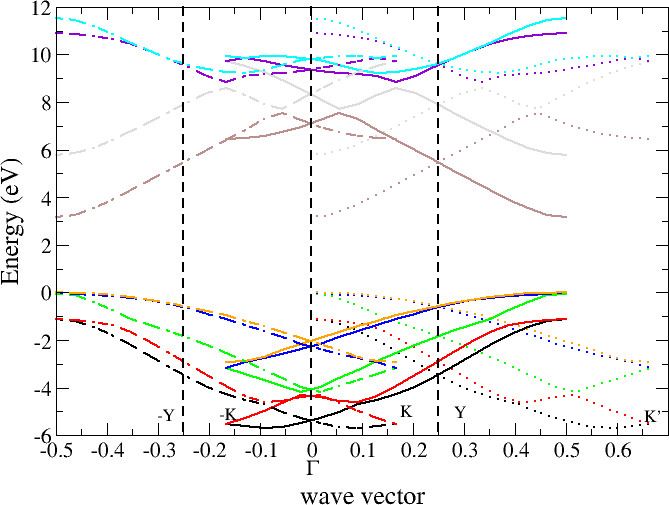}
        \caption{Folding of GaN bands along wurtzite $\Gamma-K$ onto $\Gamma-Y$ line of
                 $Pna2_1$. Original wurtzite bands(dotted), folded once about $Y$ (solid
                 lines) and folded once more about $\Gamma$, (dashed lines).\label{figGK}}
    \end{figure}

  The band structure of GaN in the wurtzite Brillouin zone is shown in Fig.~\ref{figgan}.
  Some further details of the band folding from the wurtzite to the orhorhombic Brillouin
  zone are given here. $K$ occurs at $1/3$ along the horizontal direction, while $K'$
  occurs at $2/3$ in units of $2\pi/a$. The folded position of $K$ after first folding
  about the $XS$ and then the $YS$ line is at $1/6$ along the horizontal axis while the
  point $K'$ will occur at $-1/6$.  In Fig. \ref{figGK} we show the folding of the bands
  along $\Gamma-K$ and $\Gamma-K'$ onto the $\Gamma-Y$ line of the $Pna2_1$. The color
  coding of the bands shows how they appear in pairs in the $\Gamma-X$ and $\Gamma-Y$
  lines.
  
\section{Details of exciton binding energy calculations}

    \begin{table*}[h]
      \caption{Calculation details of exciton gap and binding energy of ZnGeN$_2$. We give
               the exciton eigenvalues with respect to $N_k$ in different directions. The
               column labeled as $E_{N_k \rightarrow \infty}$ gives the exciton band gap
               obtained by the exponential fit to $N_k\rightarrow\infty$, and the column
               labeled as $E_{1/N_k \rightarrow 0}$ gives the gap values obtained by the
               linear fit to $1/N_k \rightarrow 0$. The column
               $(\varepsilon^2_{\infty}$/${\varepsilon^2_0})$ is given again for the sake
               of completeness. Finally, the last two columns give the exciton binding
               energies corresponding to different values of exciton gaps obtained from
               either fit choices. The first three rows are for a, b and c directions when
               we use $N_v=3$ and $N_c=1$. The second half is when we use $N_v=24$ and
               $N_v=16$.}
        \begin{ruledtabular}
        \begin{tabular}{l|c|c|c|c|c|c|c|c|c|c|c|c}
                & 3     & 4     & 5     & 6     & 8     & 10    & $E_{N_k \rightarrow \infty}$ & $E_{1/N_k \rightarrow 0}$ & $E_g^{QP}$ &  $(\varepsilon^2_{\infty}$/${\varepsilon^2_0})$ & BE$^{\infty}$ (eV) & BE$^0$ (eV) \\
          \hline          
          $a$   & 3.113 & 3.237 & 3.316 & 3.368 & 3.432 & 3.469 & 3.495                        & 3.509                     & 3.641      &  0.32                                           & 0.047              & 0.042       \\
          $b$   & 3.182 & 3.305 & 3.383 & 3.435 & 3.499 & 3.537 & 3.561                        & 3.576                     & 3.705      &  0.31                                           & 0.045              & 0.040       \\
          $c$   & 3.212 & 3.337 & 3.416 & 3.469 & 3.533 & 3.570 & 3.593                        & 3.609                     & 3.742      &  0.29                                           & 0.043              & 0.039       \\
          \hline                                                                                                                                                                                               
          $a$   & 3.101 & 3.226 & 3.305 & 3.358 &       &       & 3.456                        & 3.431                     &            &                                                 & 0.059              & 0.067       \\
          $b$   & 3.169 & 3.293 & 3.372 & 3.426 &       &       & 3.523                        & 3.498                     &            &                                                 & 0.056              & 0.064       \\
          $c$   & 3.201 & 3.326 & 3.406 & 3.459 &       &       & 3.556                        & 3.532                     &            &                                                 & 0.054              & 0.061       \\
        \end{tabular}
      \end{ruledtabular}\label{tabexzgn}
    \end{table*}

    \begin{table*}[t]
      \caption{Calculation details of exciton gap and binding energy of GaN. We give the
               exciton eigenvalues with respect to $N_k$ in different directions. The
               column descriptions are the same as Table \ref{tabexzgn}.The first two rows
               are for a and c directions when we use $N_v=3$ and $N_c=1$. The second half
               is when we use $N_v=6$ and $N_v=10$.}
      \begin{ruledtabular}
        \begin{tabular}{l|c|c|c|c|c|c|c|c|c|c}
              & 3     & 4     & 6     & 8     & $E_{N_k \rightarrow \infty}$ & $E_{1/N_k \rightarrow 0}$ & $E_g^{QP}$ &  $(\varepsilon^2_{\infty}$/${\varepsilon^2_0})$ & BE$^{\infty}$ (eV) & BE$^0$ (eV) \\
          \hline                                                                                                                                                                             
          $a$ & 2.741 & 2.863 & 2.995 & 3.061 & 3.114                        & 3.109                     & 3.261      &  0.28                                           & 0.041              & 0.043       \\
          $c$ & 2.786 & 2.910 & 3.042 & 3.108 & 3.158                        & 3.156                     & 3.310      &  0.32                                           & 0.049              & 0.049       \\
          \hline                                                                                                                                                                                             
          $a$ & 2.738 & 2.861 & 2.994 &       & 3.091                        & 3.050                     &            &                                                 & 0.048              & 0.059       \\
          $c$ & 2.784 & 2.908 & 3.040 &       & 3.138                        & 3.096                     &            &                                                 & 0.055              & 0.069       \\
        \end{tabular}
      \end{ruledtabular}\label{tabexgan}
    \end{table*}
  In this section, the extrapolation of the exciton gap to infinite number of k-points is
  discussed. First, the results differ slightly depending on the number of energy bands
  included. The fewer bands we included, the larger we can afford to make the number of
  k-points in the mesh but we can see from Table \ref{tabexzgn} that using more bands
  shifts the results to lower exciton gap and hence higher exciton binding energy by about
  0.01 eV. We observe a similar increase in exciton binding energies for GaN.

    \begin{figure}
      \includegraphics[width=8cm]{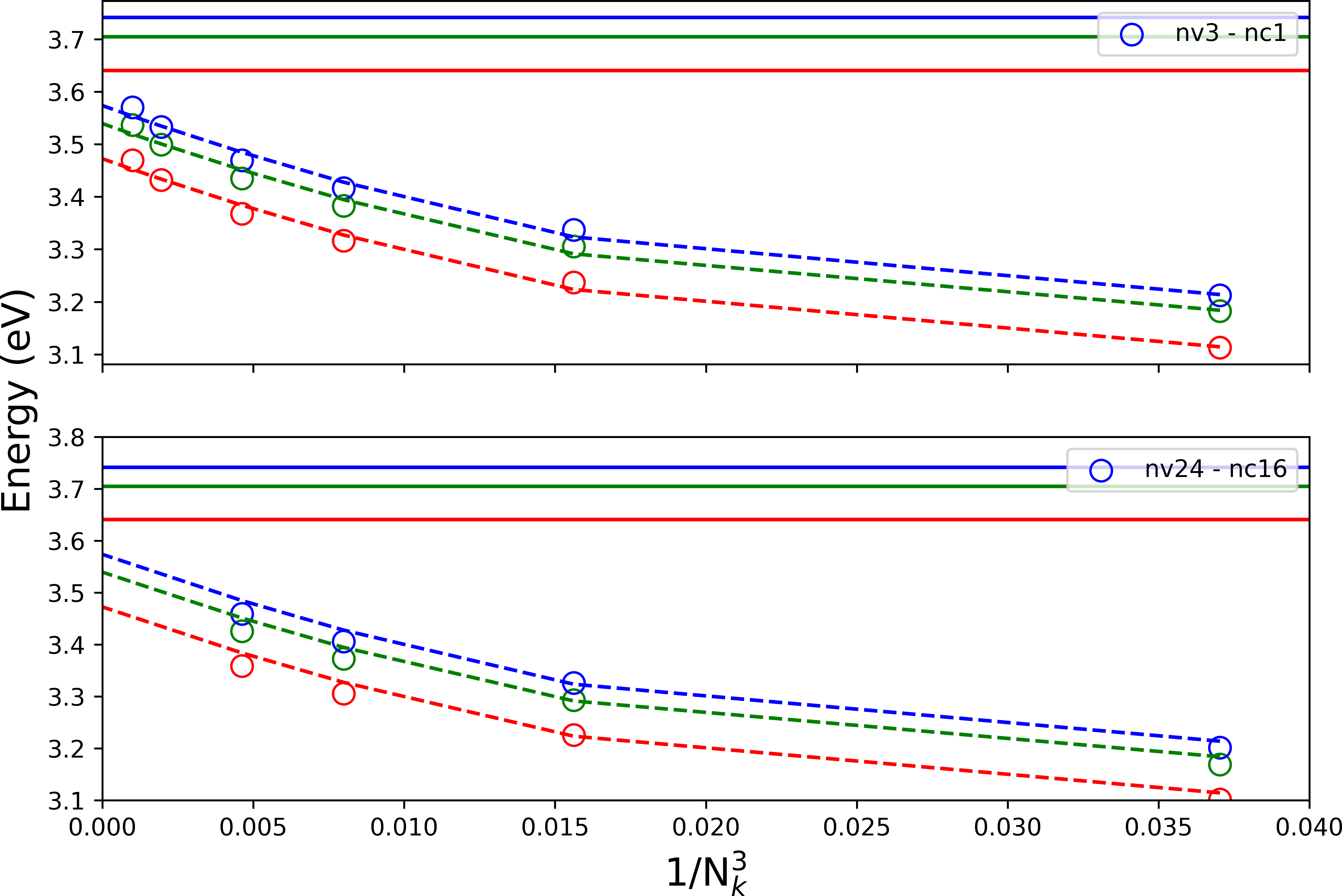}
        \caption{Exciton levels (circles) for three lowest excitons with quadratic
                 interpolation(dashed lines) and quasiparticle gaps (solid lines) for
                 three polarizatons, red (a-polarization), green (b-polarization) and blue
                 (c-polarization), as function of $1/N_k^3$. \label{figinverseNk}}
    \end{figure}
  
\begin{figure}
      \includegraphics[width=8cm]{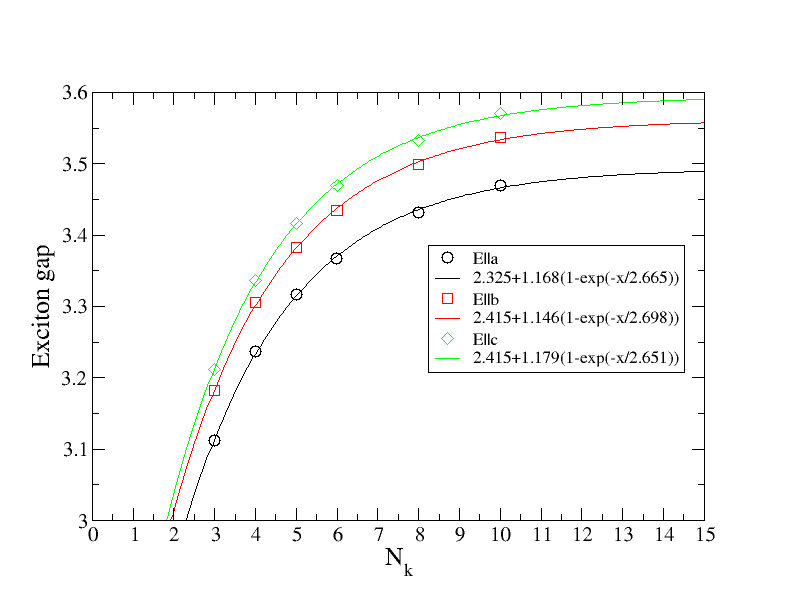}
        \caption{Exponential fits to the exciton levels as funciton of $N_k$.
                 \label{figexpon}}
    \end{figure}

  We use two different extrapolation methods in order to get the exciton gaps and binding
  energies at $N_k\rightarrow\infty$. Fig. \ref{figinverseNk} shows the  exciton levels as
  function of $1/N_k^3$ with a quadratic extrapolation. However, the smallest $1/N_k$ are
  the more relevant ones, so we can better just linearly extrapolate the two points with
  largest $N_k$. So,  the first method we use is a linear fit where we plot the exciton
  gaps with respect to $1/N_k^3$ and use the results from only the densest two
  {\bf k}-meshes available for a specific calculation. The results of this fit are
  presented under the column $E_{1/N_k \rightarrow 0}$ in Table \ref{tabexzgn} and
  \ref{tabexgan}. The second method is a fit to an exponential model function:
    \begin{eqnarray}
      E_{ex} = A + B(1 - e^{-{N_k/C}})
    \end{eqnarray}
  where we include all data points and presented under the column
  $E_{N_k \rightarrow \infty}$ in the tables. For $N_k\rightarrow\infty$ this leads to
  $A+B$ as the extrapoalted value of the exciton level. The results of this exponential
  fitting are shown in Fig.~\ref{figexpon}. In Table~\ref{tabexzgn} we can see that
  attempting to extrapolate for a larger number of bands but fewer k-points leads to
  overestimated exciton binding energies. A more accurate value is obtained by using fewer
  bands and more k-points. The uncertainty of about 0.01 eV quoted on the exciton binding
  energies is estimated from the differences between different extrapolation methods.

\section{Energy difference between band pairs and the dielectric function}
\begin{figure}
      \includegraphics[width=8cm]{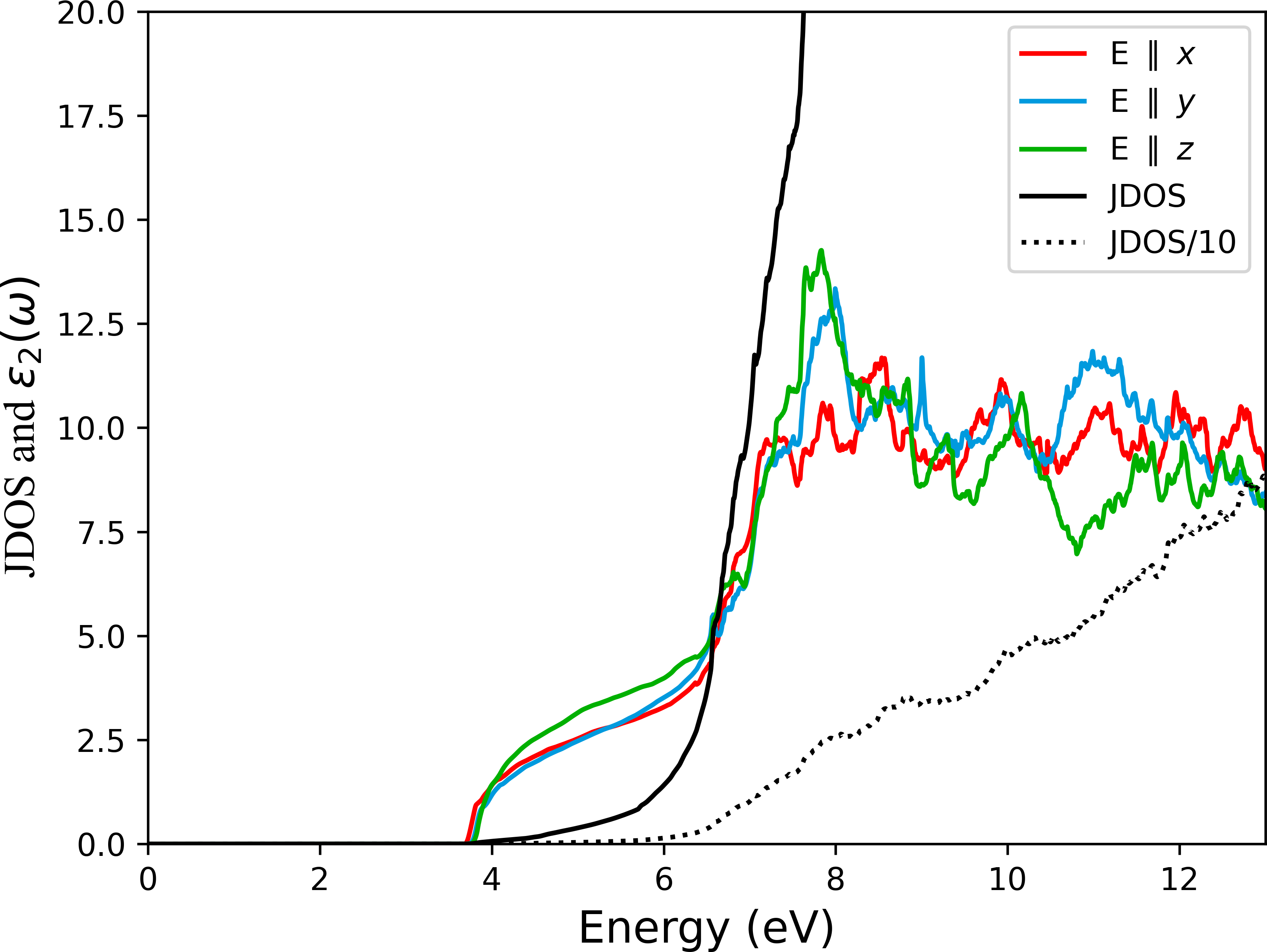}
        \caption{JDOS compared to $\varepsilon_2(\omega)$ for ZnGeN$_2$. The JDOS is in
                 units of states/eV per cell, while the $\varepsilon_2$ are dimensionless.
                 \label{figjdos}}
    \end{figure}

  In Fig.~\ref{figjdos} we compare the JDOS of ZnGeN$_2$ with the $\varepsilon_2(\omega)$
  for three polarizations.  We can see that there is some relation between the two but the
  JDOS increases much faster than the $\varepsilon_2(\omega)$. The peaks in
  $\varepsilon_2(\omega)$ rather corresponds to weaker shoulder-like features in the JDOS
  which shows oscillations on top of an overall linearly increasing background. The matrix
  elements thus play a crucial role. To explain the slow initial onset of the JDOS, which
  appears to be proportional to $\omega^2$, rather than $\omega^{1/2}$, as expected for
  parabolic bands. We note that after the initial parabolic onset of the lowest conduction
  band, it becomes linear up to about 6.5 eV. Considering that the top few valence bands
  are relatively flat, one expects  the JDOS to be essentially proportional to the
  conduction band DOS and for a linear band dispersion, one then indeed expects a
  $\omega^2$-like behavior of the DOS. This is because the number of states below a
  certain energy $\omega$ goes as $k^3$, hence as $\omega^3$ if $\omega\propto k$. The
  DOS, which is the energy derivative then goes as $\omega^2$ and is concave up. The
  $\varepsilon_2(\omega)$ including the matrix element variation, instead becomes linear
  in this range after the initial $\sqrt{\omega}$ onset. This is because the matrix
  elements strongly decrease with increasing energy, hence the $\varepsilon_2(\omega)$
  rises slower then the JDOS. 

\begin{figure*}[b]
      \begin{subfigure}[b]{0.25\textwidth}
        \hspace{-2cm}
          \includegraphics[width=6cm]{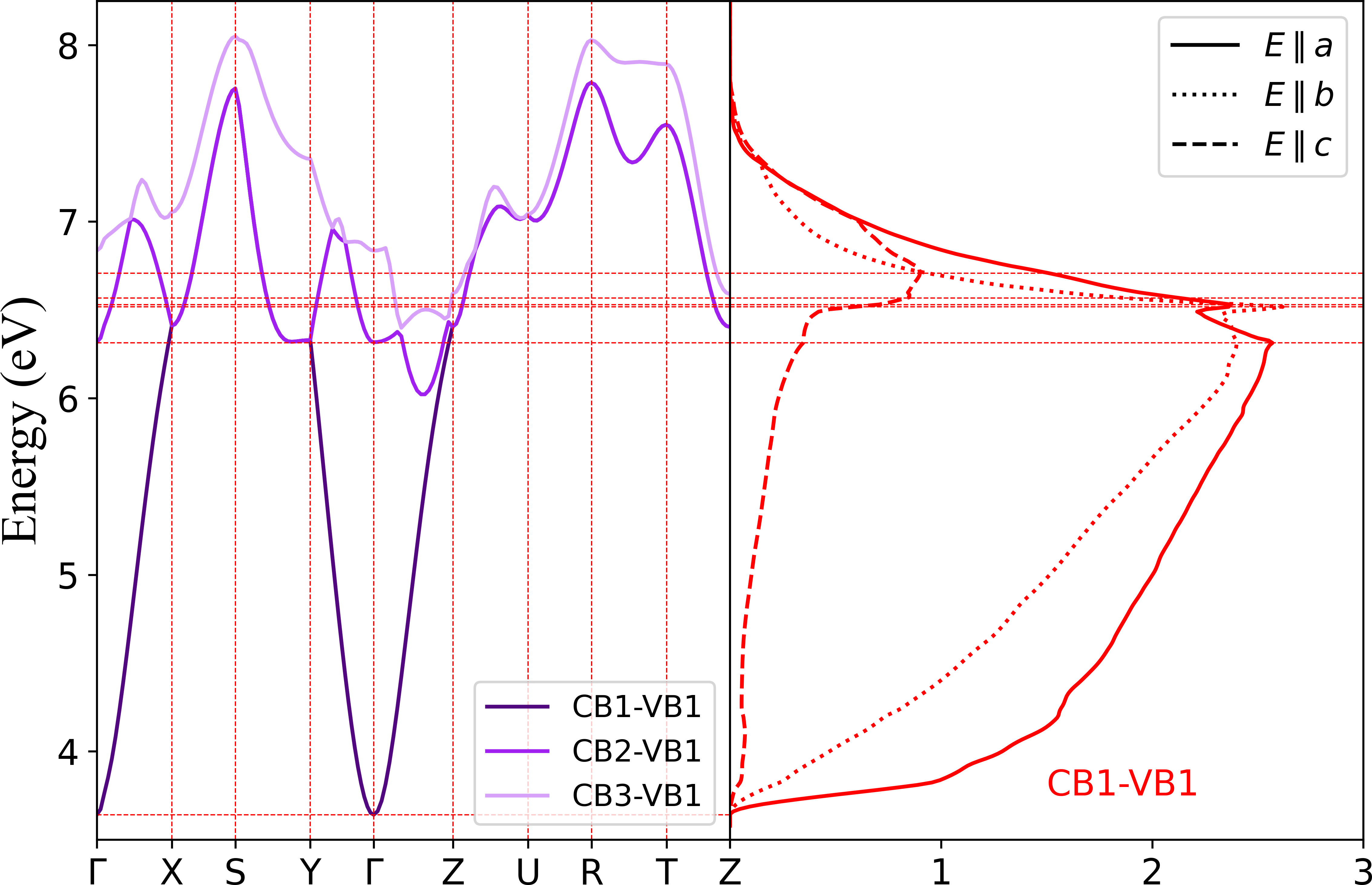}
      \end{subfigure}
        \hspace{1.5cm}
      \begin{subfigure}[b]{0.25\textwidth}
        \hspace{-2cm}
          \includegraphics[width=6cm]{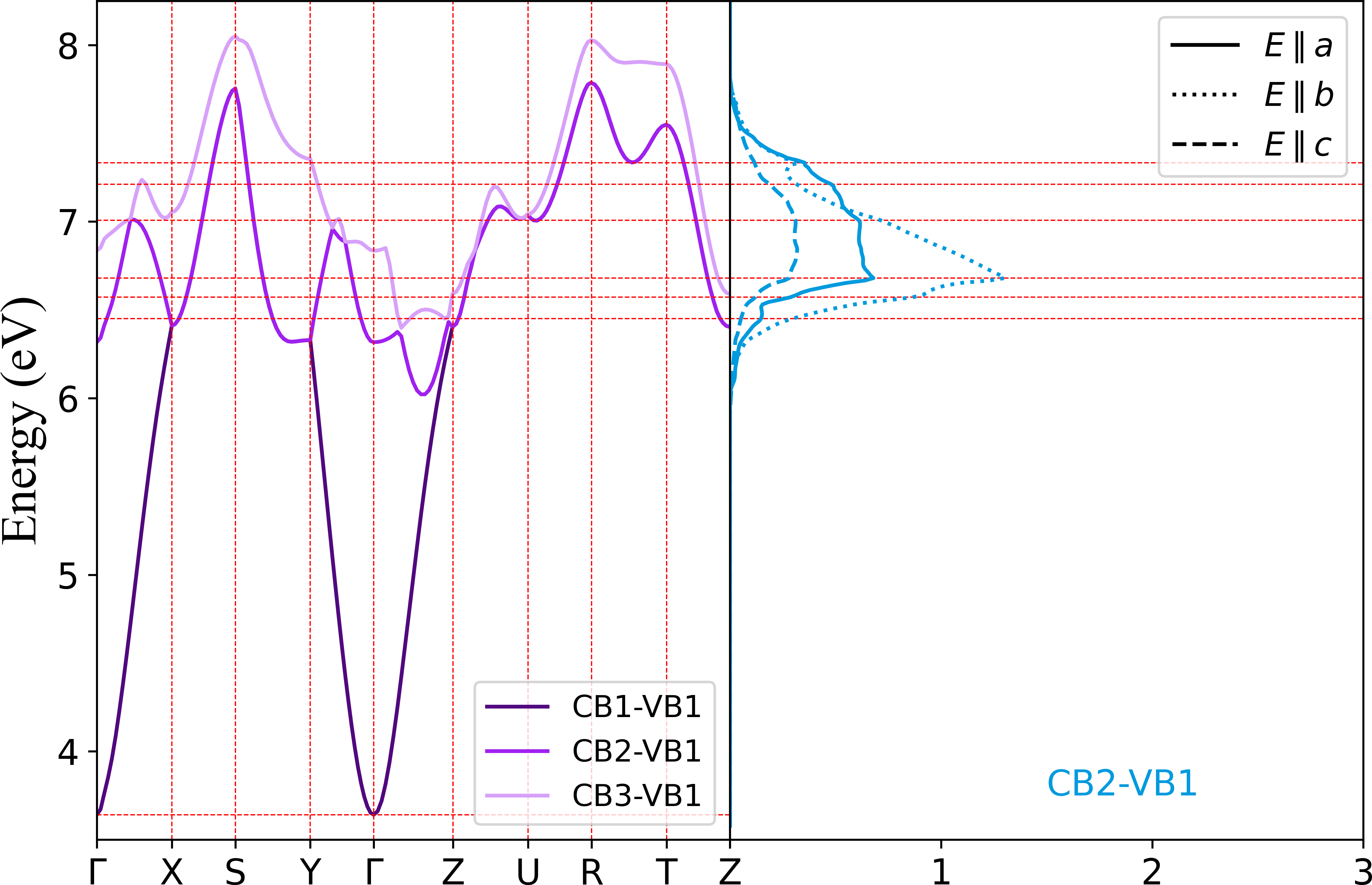}
      \end{subfigure}
      \begin{subfigure}[b]{0.25\textwidth}
          \includegraphics[width=6cm]{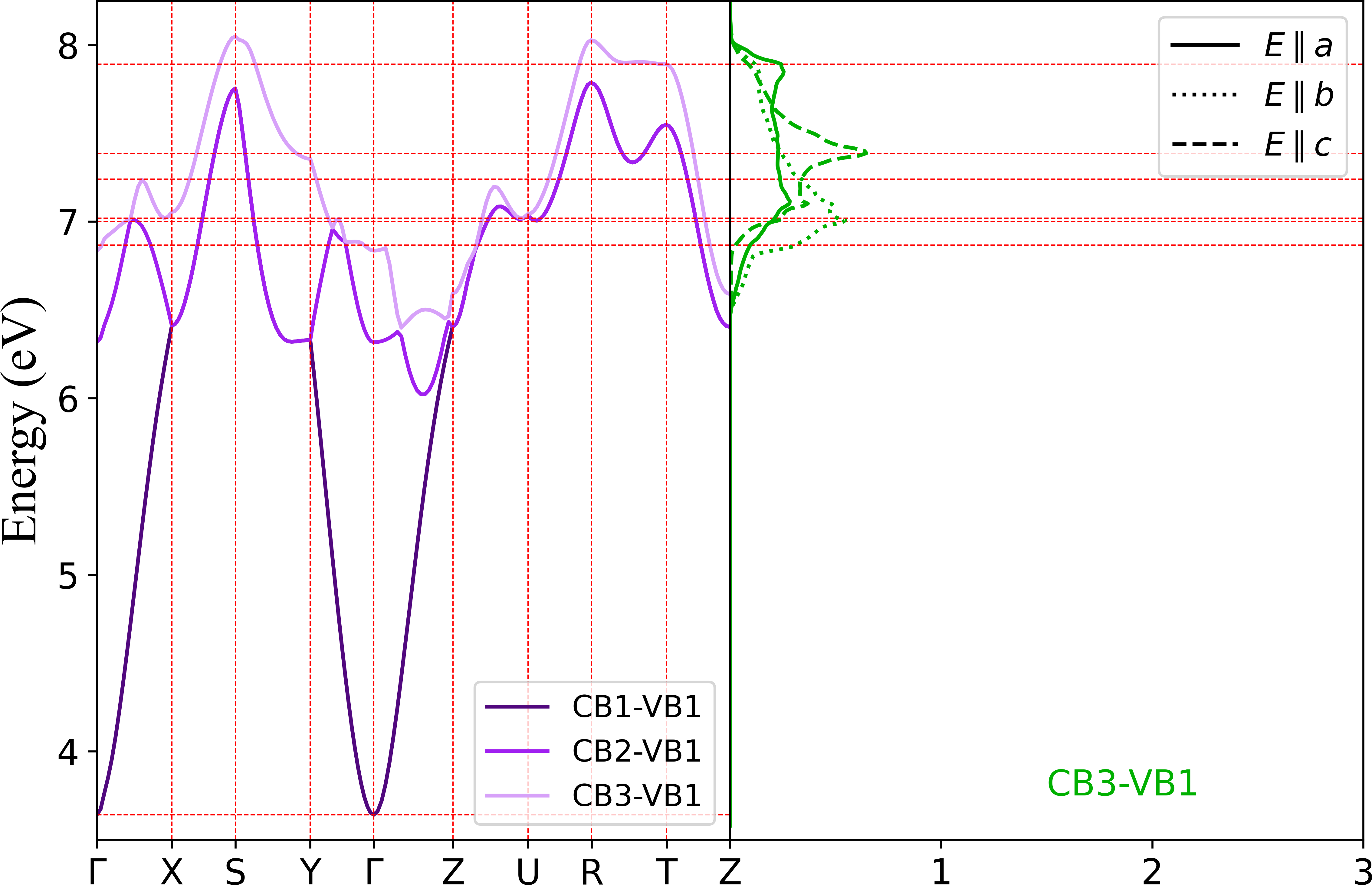}
      \end{subfigure}\\\vspace{2mm}
      \begin{subfigure}[b]{0.25\textwidth}
        \hspace{-2cm}
          \includegraphics[width=6cm]{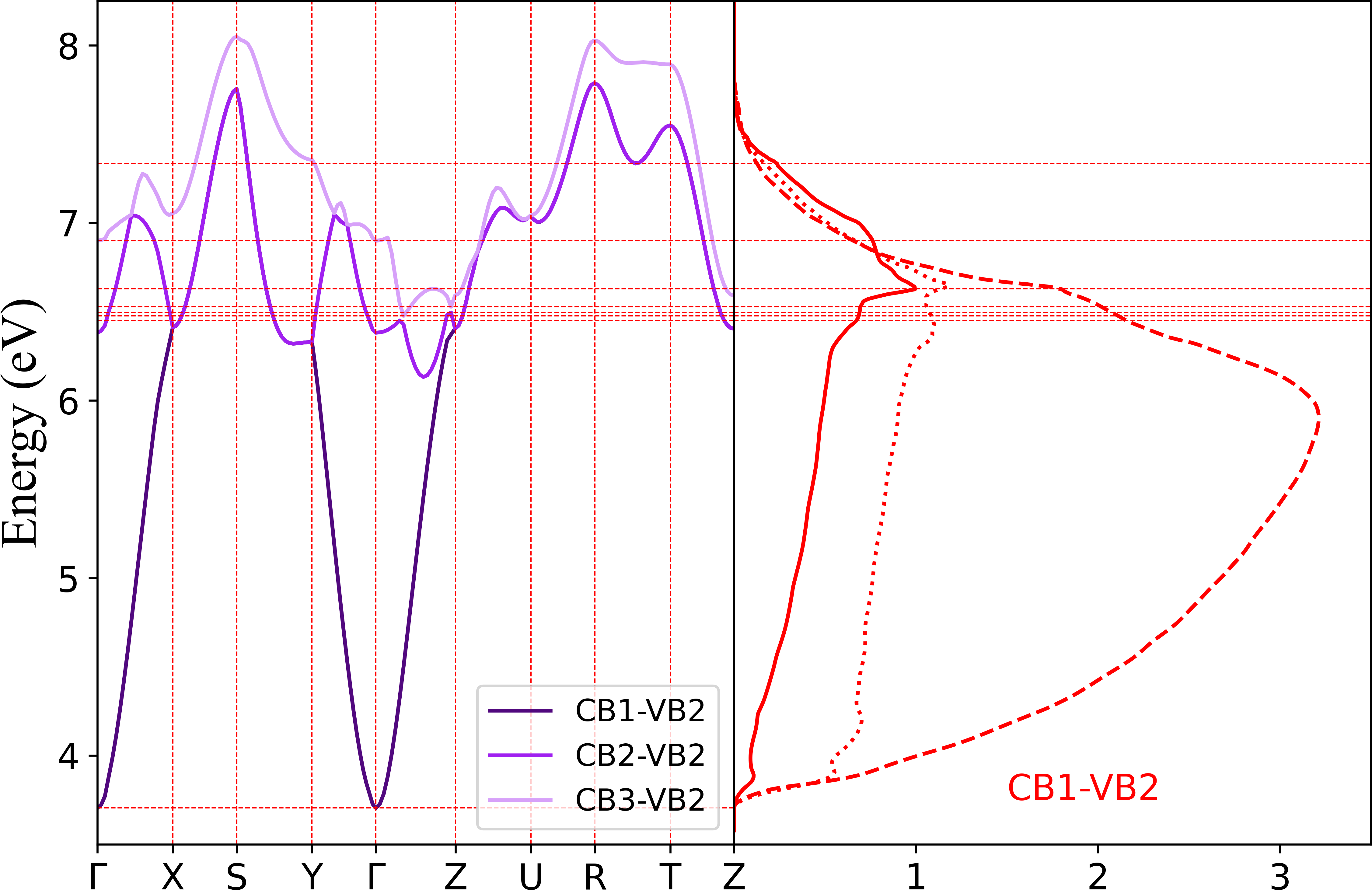}
      \end{subfigure}
        \hspace{1.5cm}
      \begin{subfigure}[b]{0.25\textwidth}
        \hspace{-2cm}
          \includegraphics[width=6cm]{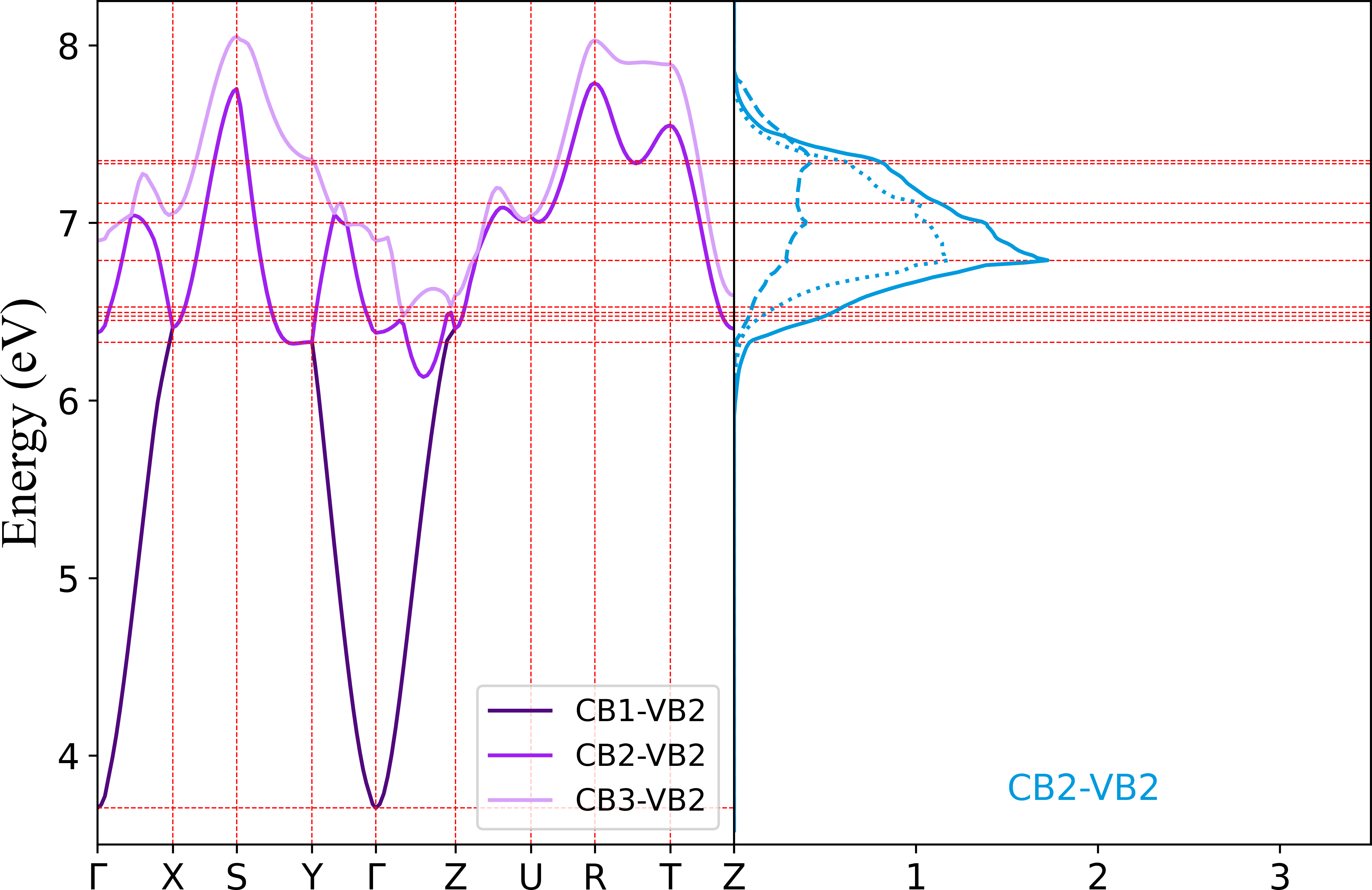}
      \end{subfigure}
      \begin{subfigure}[b]{0.25\textwidth}
          \includegraphics[width=6cm]{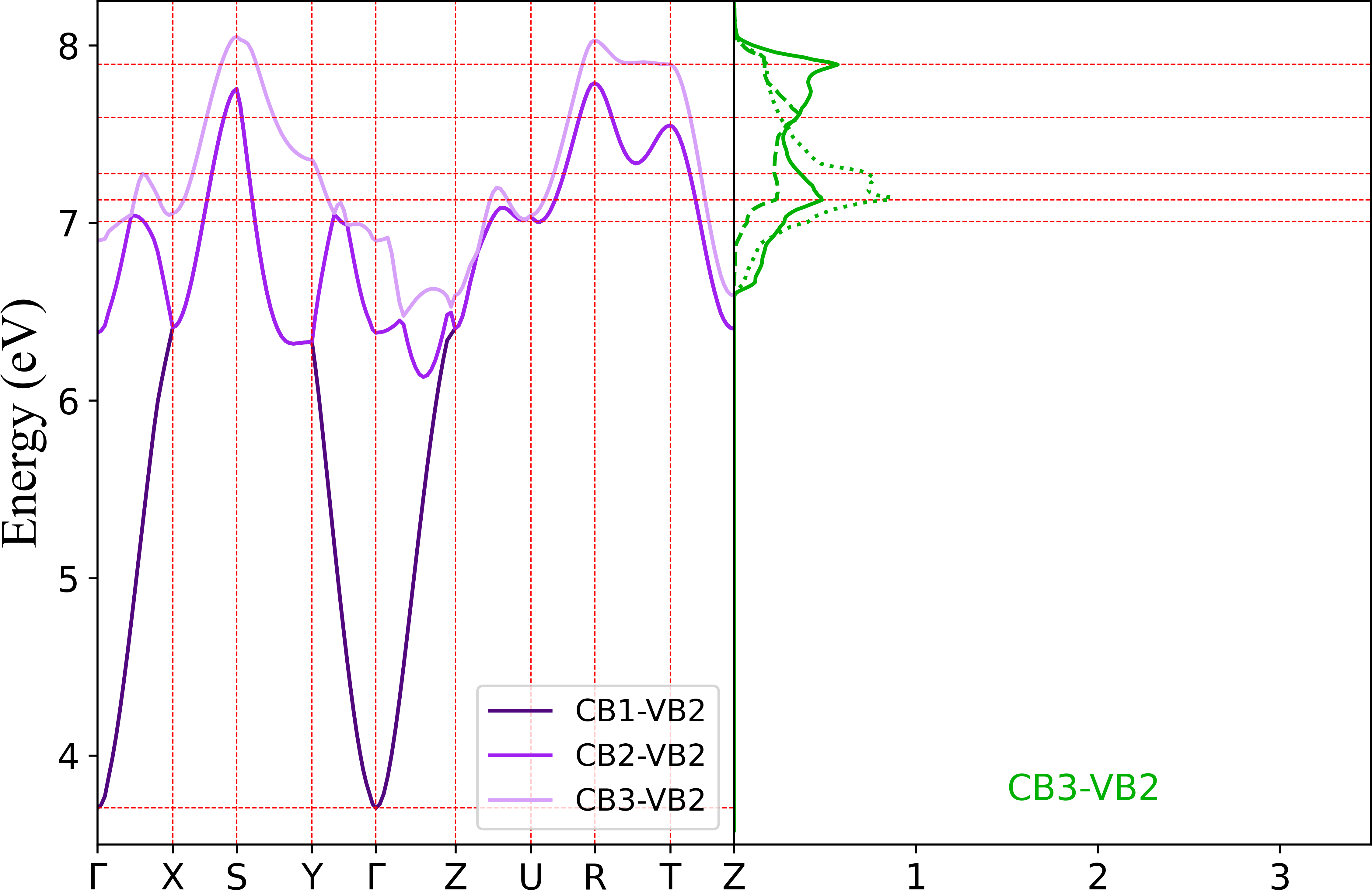}
      \end{subfigure}\\\vspace{2mm}
      \begin{subfigure}[b]{0.25\textwidth}
        \hspace{-2cm}
          \includegraphics[width=6cm]{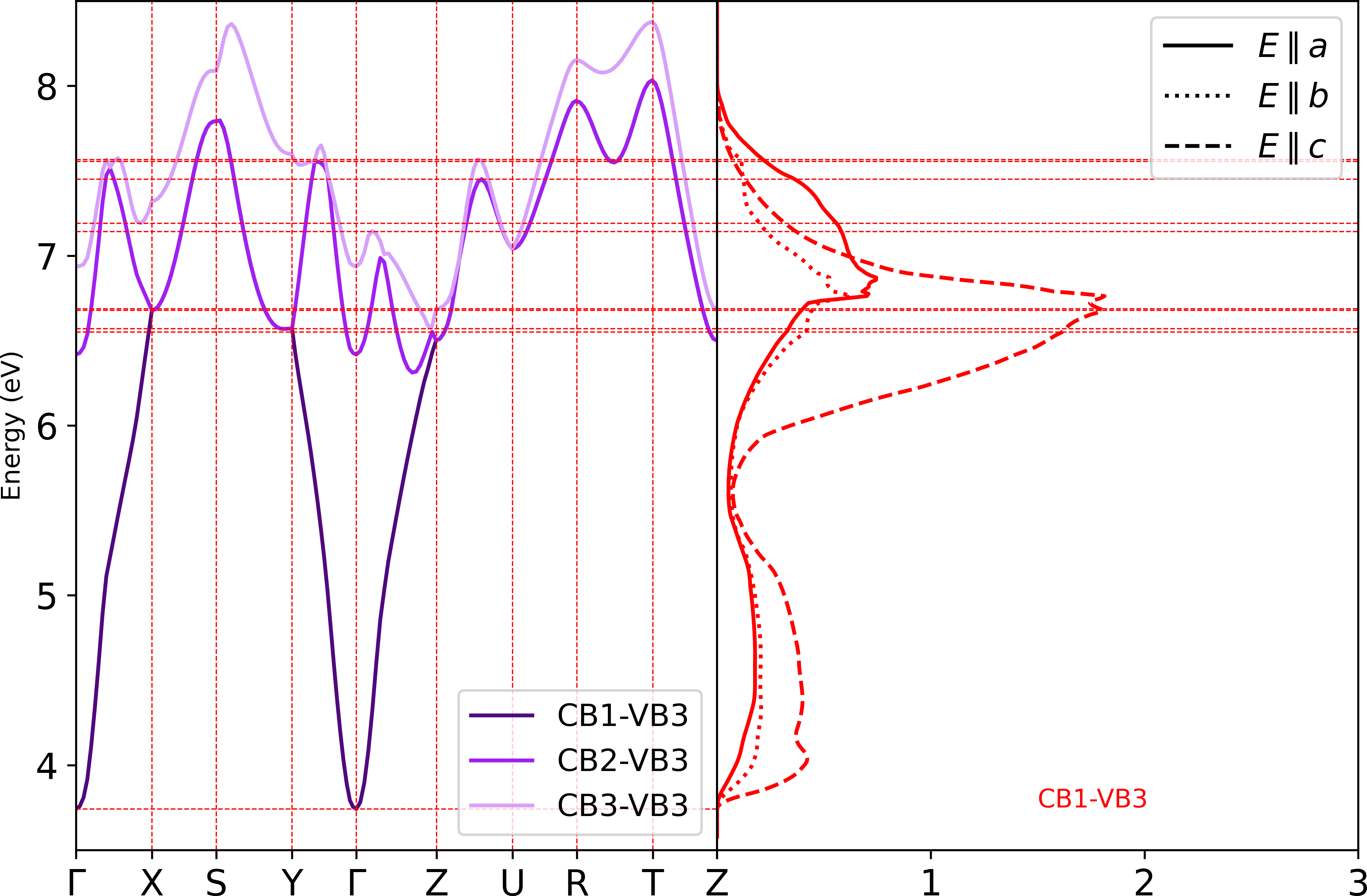}
      \end{subfigure}
        \hspace{1.5cm}
      \begin{subfigure}[b]{0.25\textwidth}
        \hspace{-2cm}
          \includegraphics[width=6cm]{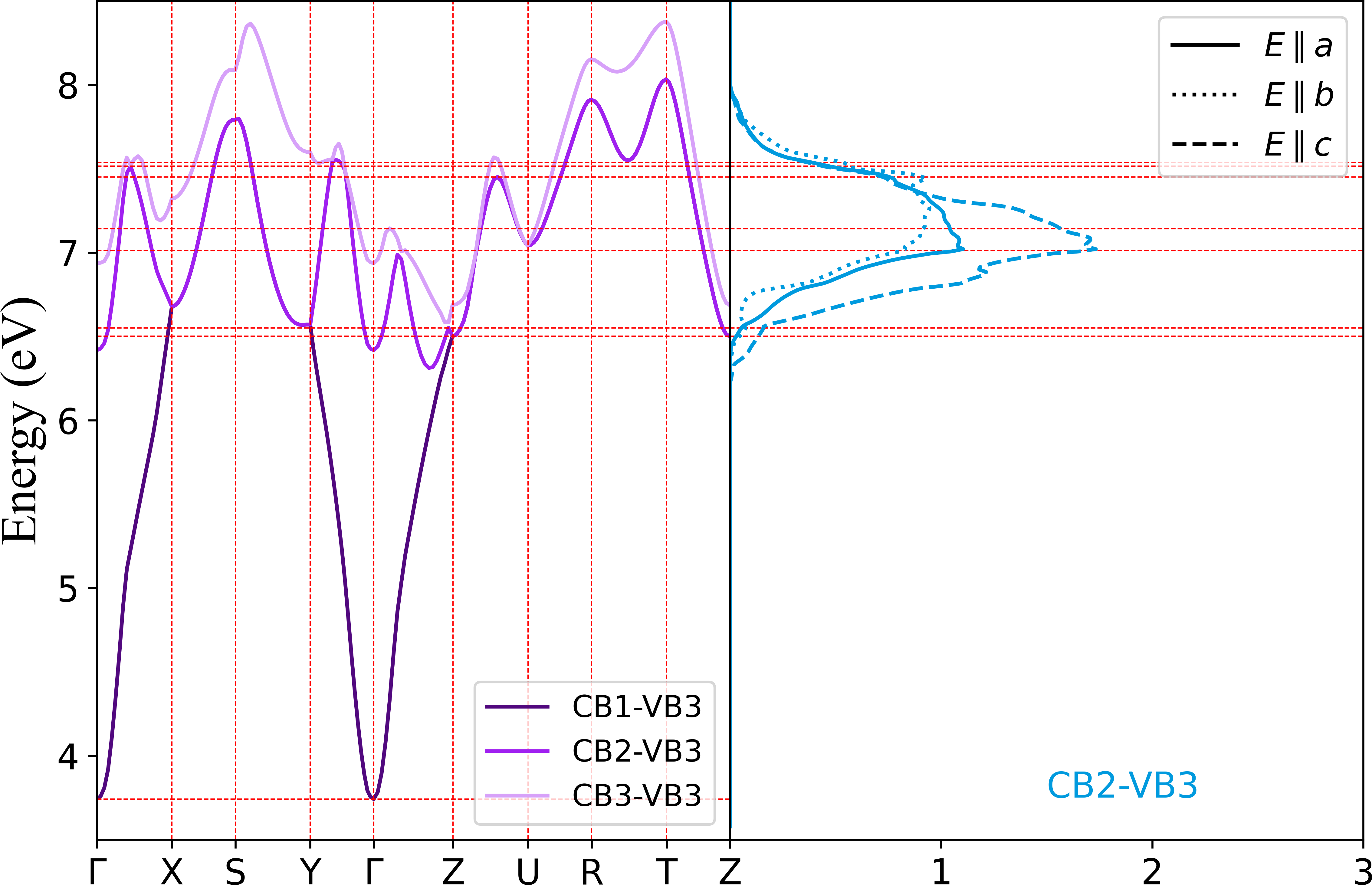}
      \end{subfigure}
      \begin{subfigure}[b]{0.25\textwidth}
          \includegraphics[width=6cm]{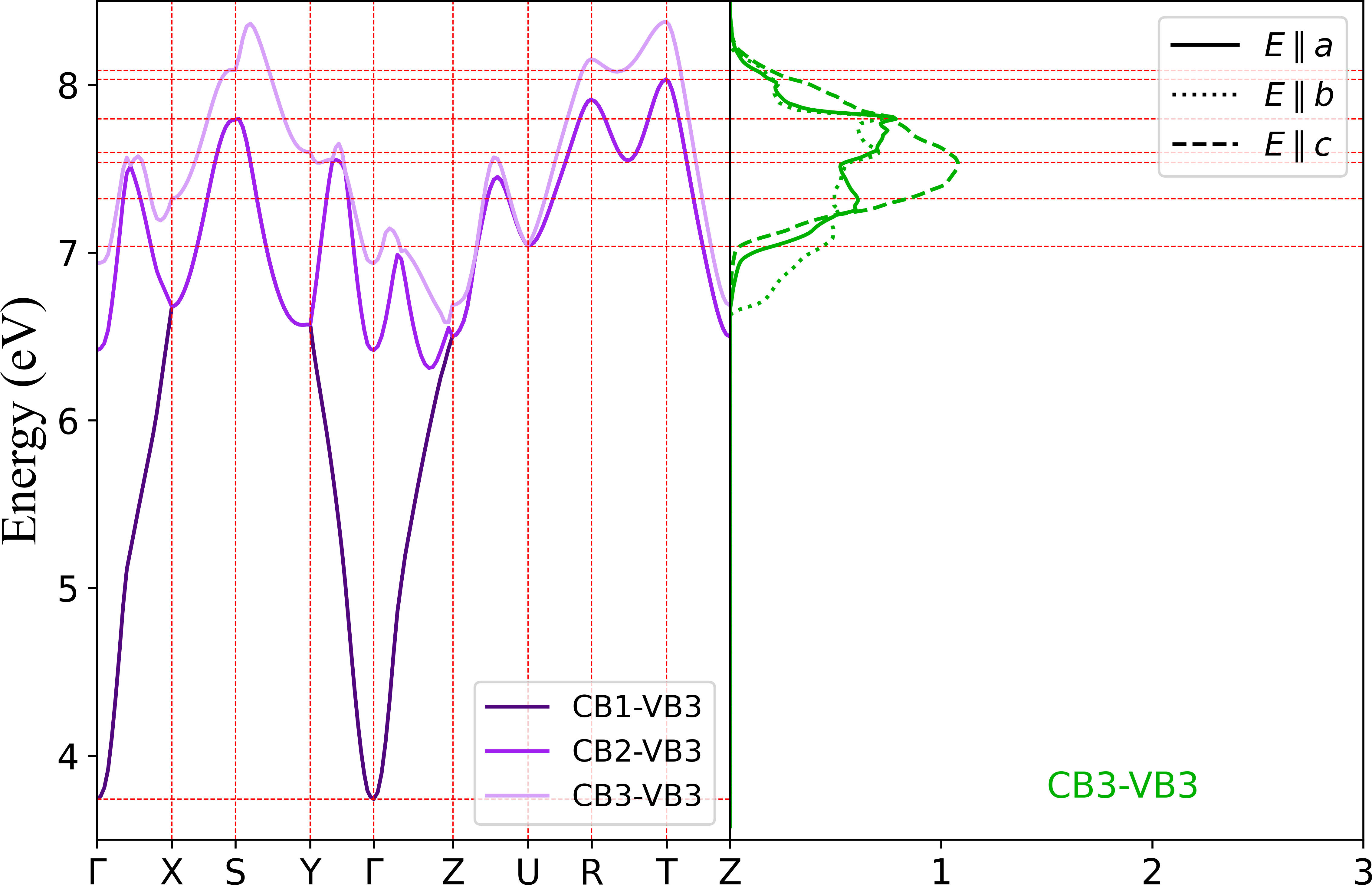}
      \end{subfigure}      
      \caption{Extended analysis of the imaginary part of the dielectric function
               $\varepsilon_2(\omega)$ versus the energy differences between the highest
               three valence bands and the lowest three conduction bands in ZnGeN$_2$.
               Three figures in the first row (CB$_{1-3}$ -- VB$_1$) are in fact given in
               the main text in a compact form in Fig.~\ref{b2bzgn} and discussed in
               detail. Additionally, here we use horizontal dashed red lines to indicate
               the energy points where a flat region in the band differences and a peak
               (or a shoulder) in $\varepsilon_2$ has been observed. The panels in the
               following two rows show the same analysis for CB$_{1-3}$ -- VB$_2$ and
               CB$_{1-3}$ -- VB$_3$.\label{complete}}
    \end{figure*}

  In the main paper, we discussed the relation between the band to band differences and
  the $\varepsilon_2(\omega)$ for the top valence band and the first three conduction
  bands. For the sake of completeness, we here also include similar plots for the VB$_2$
  and VB$_3$ energy differences with the lowest three conduction bands. These are shown in
  Fig.~\ref{complete}. Horizontal lines are used to indicate the relation between various
  features in the partial $\varepsilon_2(\omega)$ contributions for different
  polarizations and the corresponding band differences in the Brillouin zone, which allows
  one to link these features, such as changes in slope and peaks to the corresponding
  {\bf k}-locations. 
    
